\documentclass[traditabstract,bibyear,a4paper,twocolumn,utf8,hideoverfull]{aa}
\usepackage{aas_macros}

\usepackage[english]{babel}
\usepackage[utf8]{inputenc}
\usepackage[T1]{fontenc}

\usepackage{amsmath}

\usepackage{amsthm}
\usepackage{amssymb}
\usepackage{multirow}
\usepackage{algorithm}
\usepackage{algorithmic}

\usepackage{mathtools}
\usepackage{gensymb}
\usepackage{graphicx}
\usepackage[colorinlistoftodos]{todonotes}

\usepackage{amsmath}
\usepackage{appendix}
\usepackage{rotating}
\usepackage{float}
\usepackage{txfonts}
\usepackage{natbib}
\usepackage{xspace}
\usepackage{url}
\usepackage{subfigure}
\usepackage{dsfont}
\usepackage{listings}
\usepackage{color}

\usepackage{overpic}
\graphicspath{{figures/}}

\DeclareMathOperator*{\argmin}{arg\,min}
\DeclareMathOperator*{\Id}{Id}
\newcommand{\cD}{\mathcal{D}}
\newcommand{\cC}{\mathcal{C}}
\newcommand{\cT}{\mathcal{T}}
\newcommand{\bLambda}{\boldsymbol{\Lambda}}

\newcommand{\R}{\mathbb{R}}
\newcommand{\N}{\mathbb{N}}
\newcommand{\Z}{\mathbb{Z}}
\newcommand{\CC}{\mathbb{C}}
\newcommand{\DD}{\mathbf{D}}

\newcommand{\Indicator}{{\mathds{1}}}

\definecolor{dkgreen}{rgb}{0,0.6,0}
\definecolor{gray}{rgb}{0.5,0.5,0.5}
\definecolor{mauve}{rgb}{0.58,0,0.82}

\theoremstyle{remark}
\newtheorem*{rem*}{Remark}
\theoremstyle{plain}

\theoremstyle{definition}

\DeclareTextFontCommand{\emb}{\bfseries\em}

\title{Learning sparse representations on the sphere}
\titlerunning{Learning on the Sphere}
\author{F.~Sureau\inst{1}, F.~Voigtlaender\inst{2,3}, M.~Wust\inst{2}, J.-L.~Starck\inst{1}, G.~Kutyniok\inst{2}}

\authorrunning{F.~Sureau et al.}

\institute{
Laboratoire AIM, CEA, CNRS,
Université Paris-Saclay, Université Paris Diderot,
Sorbonne Paris Cité,
F-91191 Gif-sur-Yvette,
France
\and
Institut für Mathematik,
Technische Universität Berlin,
10623 Berlin,
Germany
\and
Lehrstuhl für Wissenschaftliches Rechnen,
Katholische Universität Eichstätt-Ingolstadt,
Ostenstraße 26,
85072 Eichstätt,
Germany
}

\abstract{Many representation systems on the sphere
have been proposed in the past,
such as spherical harmonics, wavelets, or curvelets.
Each of these data representations is designed to extract a specific
set of features, and choosing the best fixed representation system
for a given scientific application is challenging. 
In this paper, we show that we can learn directly
a representation system from given data on the sphere.
We propose two new adaptive approaches:
the first is a (potentially multi-scale) patch-based
dictionary learning approach, and the second consists in selecting
a representation among a parametrized family of representations,
the $\alpha$-shearlets.
We investigate their relative performance to represent and denoise complex
structures on different astrophysical data sets on the sphere.
}


\begin{document}

\maketitle

\section{Introduction}

Wavelets on the sphere \citep{starck2015} are now standard tools in astronomy
and have been widely used for purposes such as
FERMI-LAT data analysis \citep{Schmitt2010,McDermott2016},
the recovery of CMB and polarized CMB maps \citep{bobin2015,bobin2016},
string detection \citep{McEwen2017},
point source removal in CMB data \citep{sureau2014},
the detection of CMB anomalies \citep{Naidoo2017,rassat2014},
or stellar turbulent convection studies \citep{Bessolaz2011}.
While wavelets are well suited for representing isotropic components
in an image, they are far from optimal for analyzing anisotropic
features such as filamentary structures.
This has motivated in the past the construction of so called multiscale
geometric decompositions such as ridgelets,
curvelets \citep{CandesDonohoCurvelets, starck:sta02_3},
bandelets \citep{PennecM05}, or shearlets \citep{cur:labatte05}.
Extensions to the sphere of ridgelets and curvelets were already presented
in \citep{starck:sta05_2,Chan2017,McEwen2015}, and also
for spherical vector field data sets in \citep{starck:2009,Leistedt2017}.

For a given data set, we therefore have the choice between many fixed
representation spaces (pixel domain, harmonics, wavelets, ridgelets,
curvelets, etc) which are also called dictionaries.
A dictionary is a set of functions, named atoms, and the data can be
represented as a linear combination of these atoms.
The dictionary can be seen as a kind of prior \citep{Beckouche2013},
and the best representation is the one
leading to the most compact representation, that is, the maximum of information
is contained in few coefficients.
For the previously mentioned fixed dictionaries,
there exist fast operators for decomposing the data into the dictionary,
and fast operators for reconstructing the image
from its coefficients in the dictionary \citep{starck2015}.

In some cases, it is not clear which dictionary is the best,
or even if the existing dictionaries are good enough
for a given scientific application.
Therefore, new strategies were devised in the Euclidean setting to construct adaptive representations.
Among them, sparse Dictionary Learning (DL) techniques
\citep{engan:mod,ksvd:elad} have been proposed to design a dictionary
directly from the data, in such a way that the data can be sparsely represented
in that dictionary.
DL has been used in astronomy for image denoising \citep{Beckouche2013},
stellar spectral classification \citep{diaz2014} and morphological galaxy
classification \citep{diaz2016}.

An alternative approach for adaptively choosing a dictionary
is to start with a large parametrized family of dictionaries,
and then to choose the parameter(s),
either based on simulations or directly from the data.
An example of such a parametrized family of dictionaries is the family of
$\alpha$-shearlets
\citep{ShearletsNotConeAdaptedFirstPaper,AlphaMolecules,StructuredBanachFrames2}.


In this paper, we propose to extend to the sphere both adaptive
representation methods, DL and {$\alpha$}-shearlets,
and we compare the performance of the two approaches.
More precisely, we are concerned with adaptive sparsifying
representation systems for data \emph{defined on the sphere}.
In Section~\ref{sect_DL}, we present our approach for performing DL on the sphere,
while Section~\ref{sect_shear} is devoted to our extension of the
$\alpha$-shearlet transform to data defined on the sphere.
We present the scenarios for our comparison of the two approaches in
Section~\ref{sect_exp}; the results of this comparison are presented
in Section~\ref{sec:results}.
Finally, we conclude the paper in Section~\ref{sect_ccl} and 
the necessary background related to $\alpha$-shearlets in the Euclidean setting
is covered in Appendix~\ref{appendix:alpha}.
.

\section{Dictionary learning on the sphere}

\label{sect_DL}

Dictionary learning techniques have been proposed in the early 2000s
\citep{Olshausen97,Engan99,Aharon06} to build adapted linear representations
that yield sparse decompositions of the signals of interest.
Contrary to fixed dictionaries, in dictionary learning the atoms are estimated
from the data (or a proxy, such as simulations or exemplars of the data),
and can therefore model more complex geometrical content, which could ultimately
result in sparser (and typically redundant) representations.
The application of DL techniques to many inverse problems in restoration,
classification, and texture modeling has provided state-of-the-art results
(see e.g.~\citet{Elad06, Mairal08a,Mairal09,Peyre09,Zhang10}).
A wide variety of  dictionary learning techniques have been proposed to process
multivariate data \citep{Mairal08a,Mairal08b};
to construct multiscale \citep{Mairal08b},
translation-invariant \citep{Jost06,Aharon08},
or hierarchical representations \citep{Jenatton11};
to estimate coupled dictionaries \citep{Rubinstein14};
or to build analysis priors \citep{Rubinstein13}.
Also, online algorithms for dictionary learning
have been considered \citep{Mairal10}.

While fixed structured representations typically have
fast direct and inverse transforms, dictionary learning techniques
become computationally intractable even for signals of moderate size.
Based on the observation that natural images exhibit nonlocal self-similarities,
this computational problem is typically overcome by performing dictionary
learning on \emph{patches} extracted from the images that one wants to model.
In this section we focus on this patch-based dictionary learning approach,
and extend it for signals living on the sphere.

\subsection{Sparse representation with patch-based dictionary learning}

Given an $n\times n=N$ image represented as a vector $\mathbf{X}\in\R^N$,
we consider square overlapping patches $\mathbf{x}_{ij}$ in $\R^Q$,
with $Q = q \times q$, where $q$ is typically small;
in fact, in the present work we will always have $q \leq 12$.
Formally,
\begin{equation}
  \mathbf{x}_{ij} = \mathbf{R}_{ij} \mathbf{X}
\end{equation}
where the matrix $\mathbf{R}_{ij}\in\R^{Q \times N}$
extracts a patch with upper left corner at position $(i,j)$.

From a training set $\cT$ of such patches
$\left\{\mathbf{x}_{ij}\right\}_{(i,j)\in\cT}$,
a dictionary with $M$ atoms
$\DD\in\R^{Q \times M}$ is then learned such that the codes
$\bLambda=\left\{\boldsymbol{\lambda}_{ij}\right\}_{(i,j)\in\cT}$
satisfying $\mathbf{x}_{ij} = \mathbf{D} \boldsymbol{\lambda}_{ij}$ are sparse.
To perform the training, one typically considers the following
following inverse problem, or one of its variants:
\begin{equation}
\label{eq:DL_generic}
  \argmin\limits_{\DD\in\cD,\bLambda\in\cC}
    \sum_{(i,j)\in \cT}
      \| \mathbf{x}_{ij}-\DD\boldsymbol{\lambda}_{ij}\|_2^2
      + \mu \cdot \|\boldsymbol{\lambda}_{ij}\|_0
\end{equation}
where $\cD$ (resp. $\cC$) is a non-empty convex set enforcing some
constraints on the dictionary $\DD$ (resp. the codes $\bLambda$),
and $\mu \cdot \|\boldsymbol{\lambda}_{ij}\|_0$ is the weighted
$\ell_0$ pseudo-norm which enforces sparsity of the codes.
To remove the \emph{scale indeterminacy} in such a minimization problem%
---that is, if $(\mathbf{D},\mathbf{\Lambda})$ is a solution, then so is
$(\alpha \mathbf{D}, \alpha^{-1} \mathbf{\Lambda})$,
at least if $\alpha \mathbf{D} \in \cD$ and $\alpha^{-1} \mathbf{\Lambda} \in \cC$---%
the set $\cD$ is typically enforcing each atom (column) of the dictionary
to belong to a unit $\ell_2$ ball, while $\cC$ can enforce constraints in the code
(e.g. non-negativity in non-negative matrix factorization).
More details can be found in \citet{starck2015}.

\subsection{Extension of patch-based dictionary learning to the sphere}

To extend patch-based dictionary learning to data defined on the sphere,
we first need to specify how to construct patches on the sphere.
We do so by introducing local charts on the sphere.
Specifically, in this work we propose to consider the HEALPix framework
\citep{gorski98, gorski05}, widely used in astronomy, to construct these charts.

\subsubsection{Defining patches on the sphere}

\begin{figure}[h]\centering\hfill
  \includegraphics[width=63.9pt]{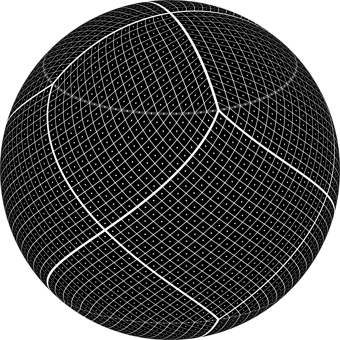}\hfill
  \includegraphics[width=127.8pt]{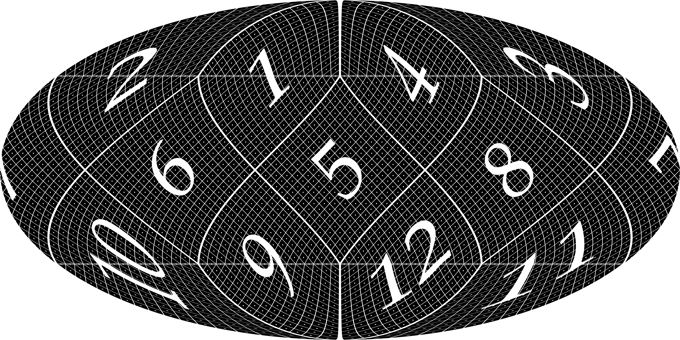}\hfill
  \caption{The HEALPix grid (visualizing $N_\text{side}=16$) in orthographic
           projection on the left and Mollweide projection on the right.
           Faint lines indicate the circles of latitude
           $\theta=\cos^{-1}(\pm\frac{2}{3})$.
           The right image also introduces the numbering of the faces,
           used in the following illustrations.}
  \label{fig:healpix_grid}
\end{figure}

HEALPix partitions the sphere into equal area pixels with curvilinear boundaries,
defined hierarchically from a set of twelve base quadrilaterals
(see Fig.~\ref{fig:healpix_grid}).
These twelve base elements (or \emph{faces}) form an atlas of the sphere,
and are further partitioned dyadically to obtain finer discretization levels.
Consequently, each of the twelve faces is typically considered as a chart
with HEALPix pixel positions mapped to a square grid in $[0,1] \times [0,1]$.

Using these charts to perform usual Euclidean patch-based dictionary learning
is straightforward, and would have the main advantage of applying
dictionary learning directly on the pixel values,
without requiring any interpolation.
This comes, however, with two drawbacks:
first, this approach introduces boundary issues even
when using overlapping patches on each face;
second, it leads to distortions for band-limited functions defined on the sphere.
While the second problem is inherent to the choice of HEALPix
as a discretization of the sphere, we can however
address the first problem in this framework:
patches can be created based on local neighbors
as defined in the HEALPix framework.
Because of the regularity of the HEALPix sampling scheme,
all pixels have eight neighbors, except for eight pixels on the sphere
that are located at the vertices in between equatorial and polar faces,
which only have seven neighbors.

\begin{figure}[h]
  \begin{center}
   \includegraphics[width=164pt]{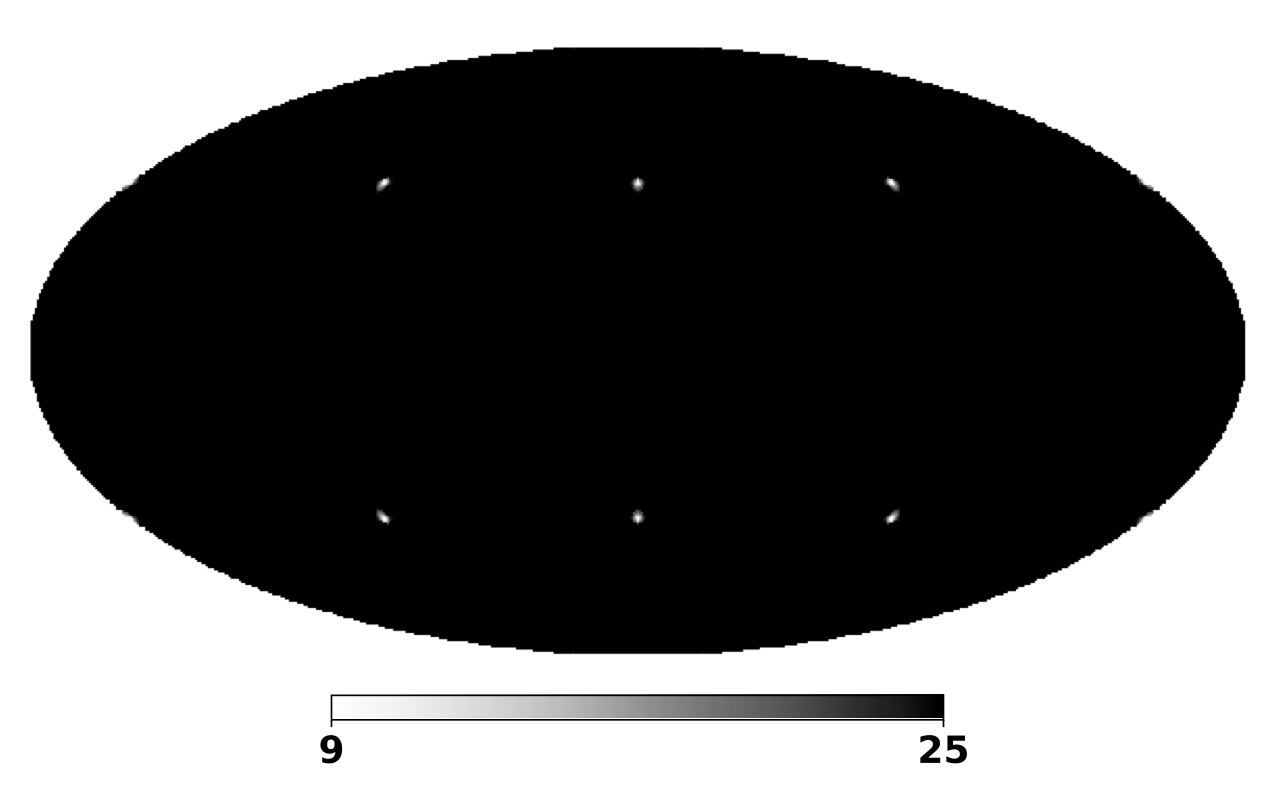}
  \end{center}
  \caption{\label{CoverMap}
           Example of our covering of the sphere with overlapping patches
           based on HEALPix neighborhoods.
           The plotted value indicates the number of overlapping patches
           including each pixel, for $N_{side}=128$ and patch width $q=5$.
           Because the patch width is usually small
           with respect to the number of pixels per face,
           only a small fraction of patches is not considered
           (in this work, $N_{side}=2048$ and the patch width is $q=8$ or $q=12$).}
\end{figure}

Provided some care is taken on defining the respective position of each
neighbour to a central pixel across the sphere, overlapping patches
can be created---even in between the twelve HEALPix faces---without
any interpolation, except at the patches crossing the specific points
on the HEALPix grid which only have seven neighbors.
Interpolation strategies to compensate for these "missing" neighbors
can be envisioned; but in this work we choose not to interpolate,
which implies that for a few pixels around these points,
we do not construct all overlapping patches.
The final covering of the map is illustrated in Fig.~\ref{CoverMap}.
Once these patches are extracted, classical dictionary learning
techniques can be used to learn a sparse adapted representation.

\subsubsection{Learning a multi-scale representation on the sphere}

Our proposed approach for dictionary learning on the sphere can be extended
to capture multiscale information as proposed in \citet{Ophir2011}:
a dictionary is learned from patches
extracted from a multiscale decomposition of the data.

At lower scales, capturing information would require to increase the patch size,
and would ultimately lead to a computational burden impossible to handle.
To capture this information without increasing the patch size,
the decomposition is subsampled.

In this work, we use the Starlet decomposition
for data on the sphere \citep{starck:sta05_2},
with one dictionary learned per wavelet scale.
Since all scales except the first one are band-limited,
subsampling can be performed without loosing information by adapting
the $N_{side}$ parameter to the maximal multipole at the level considered
(typically dyadically decreasing, as illustrated in Table \ref{tab:DLDustParams}).

\begin{table}
  \caption{Parameters used for learning the multiscale dictionary
           for thermal dust data.
           For each Starlet scale, the maximal multipole $\ell_{max}$,
           the $N_{side}$ parameter,
           the number of patches,
           their width $q$,
           the number of atoms $M^{(s)}$,
           the maximal sparsity $K^{(s)}$,
           and the number iterations $N_{it}$ are displayed.
  \label{tab:DLDustParams}}
  \begin{center}
    \begin{tabular}{|l|l|l|l|l|l|l|l|}
      \hline
      Scale & $\ell_{max}$ & $N_{side}$ & NPatch & $q$ & $M^{(s)}$ & $K^{(s)}$ & $N_{it}$\\
      \hline
      0 & n.a. & 2048 & 200k & 12 & 256  & 10 & 100\\
      \hline
      1 & 1024 & 512 & 50k   & 12 & 256 & 20 & 100\\
      \hline
      2 & 512 & 256 & 25k   & 12 & 256 & 30 & 100\\
      \hline
    \end{tabular}
  \end{center}
\end{table}

The resulting minimization problem for the
multiscale dictionary learning problem reads:
\begin{equation}
  \label{eq:DL_multiscale_generic}
  \argmin\limits_{\substack{\{\DD^{(s)}\}_{s=0..S}\in\cD,\\
                            \{\bLambda^{(s)}\}_{s=0..S}\in\cC}}
    \quad \sum^{S}_{s=0}
              \sum_{(i,j)\in\cT^{(s)}}
                \|
                  \mathbf{R}_{ij} \mathcal{W}^{(s)} \mathbf{X}
                  - \DD^{(s)}\boldsymbol{\lambda}^{(s)}_{ij}
                \|_2^2
                + \mu^{(s)} \cdot \| \boldsymbol{\lambda}^{(s)}_{ij} \|_0
\end{equation}
where $\mathbf{X}$ is the signal on the sphere, $\mathcal{W}^{(s)}$ extracts the scale $s$
 of the wavelet transform on the sphere according to the $N_{side}$ chosen for that scale,
$\mathbf{R}_{ij}$ is now extracting patches according to neighbors on the sphere for the patch
indexed by $(i,j)$ at scale $s$ in training set $\cT^{(s)}$,
and $S$ is the total number of wavelet scales.
For each scale $s = 0, \dots, S$, a dictionary $\DD^{(s)}$ is therefore learned,
giving coefficients $\boldsymbol{\lambda}^{(s)}_{ij}$ collected in $\bLambda^{(s)}$;
the hyperparameter $\mu^{(s)}$ is also allowed to change with the scale.

Because the cost function is separable per scale,
the minimization problem \eqref{eq:DL_multiscale_generic}
is equivalent to solving $S+1$ dictionary learning sub-problems
associated to each wavelet scale.

\subsection{Our algorithm for patch-based dictionary learning on the sphere}

In the training phase, the joint nonconvex problems described in
Eqs.~\eqref{eq:DL_generic}-\eqref{eq:DL_multiscale_generic}
are typically handled by alternating sparse coding steps
and dictionary update steps.

Here, a sparse coding step means that one minimizes
Eq.~\eqref{eq:DL_generic} (resp. Eq.~\eqref{eq:DL_multiscale_generic})
with respect to $\bLambda$ (resp. $\bLambda^{(s)}$),
with a fixed previously estimated dictionary.
Similarly, a dictionary update step means that one minimizes
Eq.~\eqref{eq:DL_generic} (resp. Eq.~\eqref{eq:DL_multiscale_generic})
with respect to $\DD$ (resp. $\DD^{(s)}$),
with the fixed previously estimated codes.
Note that both sub-problems can be minimized with standard algorithms.
In this work, we will use the classical dictionary learning technique
K-SVD \citep{Aharon06} with Orthogonal Matching Pursuit (OMP)
\citep{Mallat93,Pati93} as a sparse coder.
For denoising applications, the sparse coding step will encompass
both a maximal sparsity level, and an approximation threshold
based on the $\ell_2$ norm of the residual,
similar to the approach in \citet{Elad06}.
This approach resulted in adapted sparse representations,
while not being sensitive to small fluctuations below the targeted
level of approximation, and in practice led to faster algorithms.

The resulting multiscale dictionary learning algorithm is described in
Algorithm~\ref{Algo:DL}, from which its variant without the multiscale
transform can be obtained for $S = 0$ and $\mathcal{W}^{(0)} = \Id$.

\begin{algorithm}
  \caption{Multiscale Dictionary Learning on the Sphere \label{Algo:DL}}
  \begin{algorithmic}[1]
  \STATE \textbf{Initialization}: For each scale $s = 0,\dots,S$,
                                  choose the number of atoms $M^{(s)}$,
                                  a maximal sparsity degree $K^{(s)}$,
                                  a maximal approximation error $\epsilon^{(s)}$.
                                  Initialize the dictionary.
                                  Choose the number of iterations $N_{it}$.
  \STATE \textbf{Patch Extraction}: For each scale $s$, extract randomly patches
                                    $\left\{\mathbf{R}_{ij} \mathcal{W}^{(s)} \mathbf{X}\right\}_{(i,j)\in\cT^{(s)}}$
                                    on the sphere.
                                    Subtract from each patch its mean value.
  \FOR[\textbf{Subproblem for scale $s$}]{$s=0$ to $S$}
  \FOR[\textbf{Main Learning Loop}]{$n=0$ to $N_{it}$}
  \FOR[\textbf{Sparse Coding}]{$(i,j)\in\cT^{(s)}$}
  \STATE Compute the sparse code $\boldsymbol{\lambda}^{(s)}_{ij}$
         using OMP with stopping criterion
         $\|
            \mathbf{R}_{ij} \mathcal{W}^{(s)} \mathbf{X}
            - \DD^{(s)}\boldsymbol{\lambda}^{(s)}_{ij}
          \|_2<\epsilon^{(s)}$
         or $\|\boldsymbol{\lambda}^{(s)}_{ij}\|_0 > K^{(s)}$
  \ENDFOR
  \STATE Update $\DD^{(s)}$ using K-SVD \citep{Aharon06} \COMMENT{\textbf{Dictionary Update}}
  \ENDFOR
  \ENDFOR
  \RETURN $\left\{\DD^{(s)}\right\}_{s=0..S}$
  \end{algorithmic}
\end{algorithm}

The first critical choice for this dictionary learning technique is to adapt
the patch size $q$ to capture information at the scale of the patch without
impacting too much the computational burden of the algorithm
($q$ is at most $12$ in this work).
The maximal sparsity degree $K^{(s)}$ and the number of atoms $M^{(s)}$
should be selected so that the dictionary leads to small approximation errors,
while being able to capture the important features with only a few atoms,
in particular for denoising applications.
The parameter $\epsilon^{(s)}$ is the level of the noise
expected in the denoising application at the considered wavelet scale,
and the number of iterations is in practice chosen sufficiently large
so that the average approximation error does not change with iterations.
Because this problem is non-convex, it is crucial to initialize the algorithm
with a meaningful dictionary;
in our case, the initial dictionary is chosen to be an
overcomplete discrete cosine transform (DCT) dictionary as in \citet{Elad06}.

\section{$\alpha$-shearlets on the sphere}

\label{sect_shear}

\subsection{Euclidean $\alpha$-shearlets}

$\alpha$-shearlets are a family of representations that generalizes
wavelets and shearlets; the family is parametrized by the
\emph{anisotropy parameter} $\alpha \in [0,1]$.
To each parameter $\alpha$ corresponds a dictionary characterized by:
\begin{itemize}
  \item atoms with a ``shape'' governed by
        $\mathrm{height} \approx \mathrm{width}^\alpha$
        (see Fig.~\ref{fig:AlphaEffect});

  \item a directional selectivity: on scale $j$, an $\alpha$-shearlet
        system can distinguish about $2^{(1-\alpha)j}$ different directions
        (see Fig.~\ref{fig:AlphaShearletFrequencyConcentration});

  \item a specific frequency support for the atoms
        (see Fig.~\ref{fig:AlphaShearletFrequencyConcentration}).
\end{itemize}

A key result \citep{StructuredBanachFrames2}
is that $\alpha$-shearlets are almost optimal for the approximation
of so-called $C^\beta$-cartoon-like functions, a model class for natural images.
More precisely, the $N$-term $\alpha$-shearlet approximation error
(that is, the smallest approximation error that can be obtained using a linear
combination of $N$ $\alpha$-shearlets) for a $C^\beta$-cartoon-like function
is decreasing at (almost) the best rate that \emph{any} dictionary can reach for
the class of such functions.
For this to hold, the anisotropy parameter $\alpha$ needs to be adapted to the
regularity $\beta$, that is, one needs to choose $\alpha = 1/\beta$.
For more details on this, we refer to Appendix~\ref{appendix:alpha}.

In general, given a certain data set, or a certain data model,
different types of $\alpha$-shearlet systems will be better adapted
to the given data than other $\alpha'$-shearlet systems.
Thus, having such a versatile, parametrized family of representation system
is valuable to adapt to a variety of signals to recover.


\subsection{Extending $\alpha$-shearlets to the sphere}

In order to define the $\alpha$-shearlet transform on the sphere,
similarly to what was discussed for the dictionary learning approach,
we need to define the charts on which the
Euclidean $\alpha$-shearlet transform will be applied.
HEALPix faces are again an obvious candidate since these \emph{base resolution pixels}
can be interpreted as squares composed of $N_\text{side}$ by $N_\text{side}$
equally spaced pixels, although their shape is contorted in different
ways on the sphere (see Fig.~\ref{fig:healpix_grid}).
We could map the sphere to these twelve square faces and then
take the $\alpha$-shearlet transform on every one of them individually.
However, as for dictionary learning, this approach to the processing of HEALPix data
(e.g. for the task of denoising) is deemed to introduce boundary artefacts
due to the disjoint nature of the partition.
An example of such artefacts can be seen in the upper-left part of
Fig.~\ref{fig:TDustArtefacts} shown in Section~\ref{sec:results}.
Note also that contrary to the patch-based dictionary learning
where the patch size remains typically small compared to a face size,
the increasing size of the $\alpha$-shearlet atoms when going to lower
scales can introduce large border effects.

In the following two subsections, we discuss two approaches for
handling this problem.

\subsubsection{The rotation-based approach}
\label{sub:PartitionOfUnity}

The first strategy to alleviate the block artefacts was proposed
for curvelets in \citet{starck:sta05_2}.
This approach relies on considering \emph{overlapping} charts
that are obtained by considering HEALPix faces after
resampling the sphere through a small number of rotations.

More precisely, for a given Euclidean $\alpha$-shearlet system,
a HEALPix face $f$, and a rotation $\mathbf{r}$,
the redundant coefficients are obtained by:
\begin{equation}
  \label{eq:rot_decomp}
  \boldsymbol{\lambda}_{\alpha,\mathbf{r},f}
  = \mathcal{S}_\alpha \left(
                         \mathbf{H}_f
                         \mathcal{R}_{\mathbf{r}}\left(
                                                   \mathbf{X}
                                                 \right)
                       \right),
\end{equation}
where $\mathcal{R}_{\mathbf{r}}$ is computing the resampled map
by a rotation $\mathbf{r}$ of the sphere,
$\mathbf{H}_{f}$ is a matrix extracting the pixels that belong to the HEALPix face $f$,
and $\mathcal{S}_\alpha$ is computing the Euclidean $\alpha$-shearlet transform
on this face.
In practice, a bilinear interpolation is performed
by the HEALPix rotation routines that are used for the resampling.

\begin{figure}[h]
  \begin{center}
    \begin{tabular}{cc}
      \includegraphics[width=75pt]{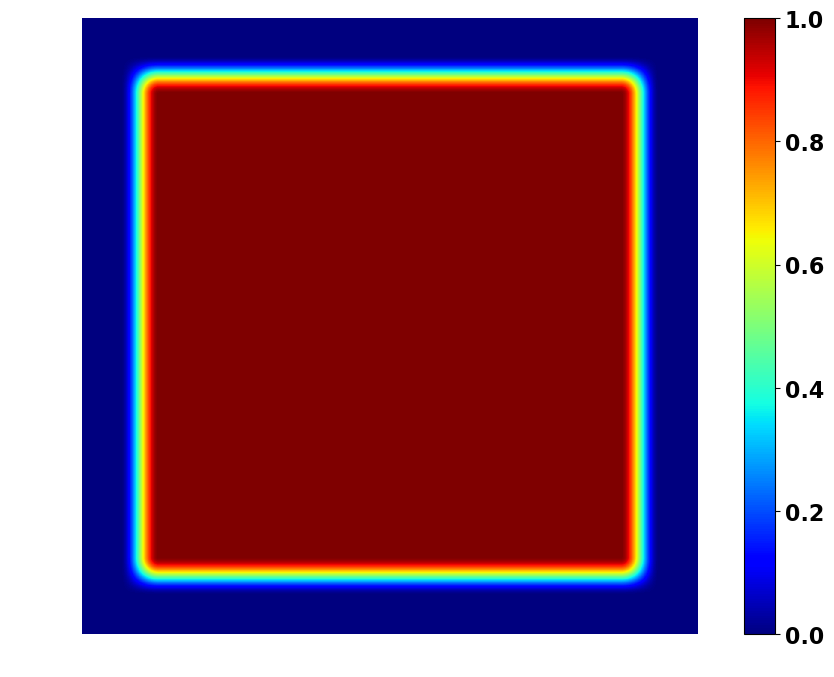} &
      \includegraphics[width=110pt]{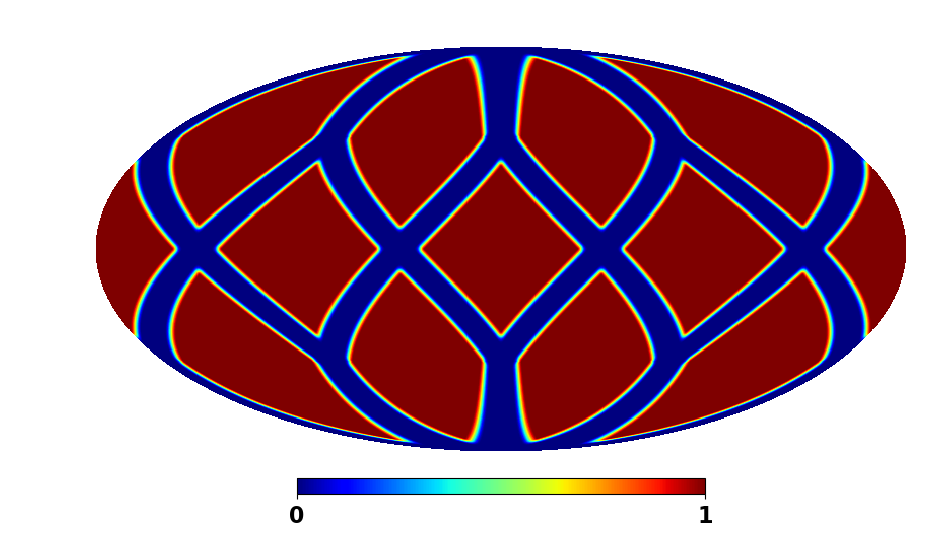} \\
      \includegraphics[width=110pt]{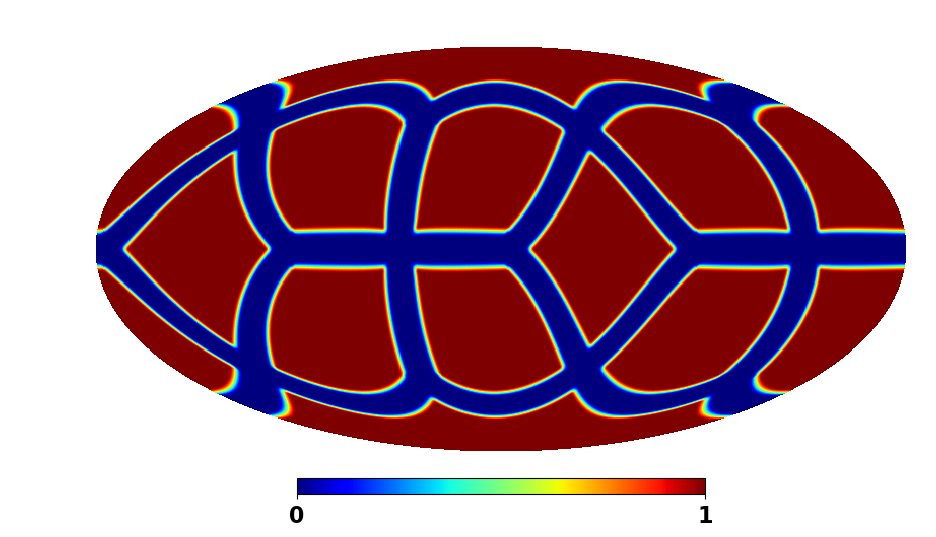} &
      \includegraphics[width=110pt]{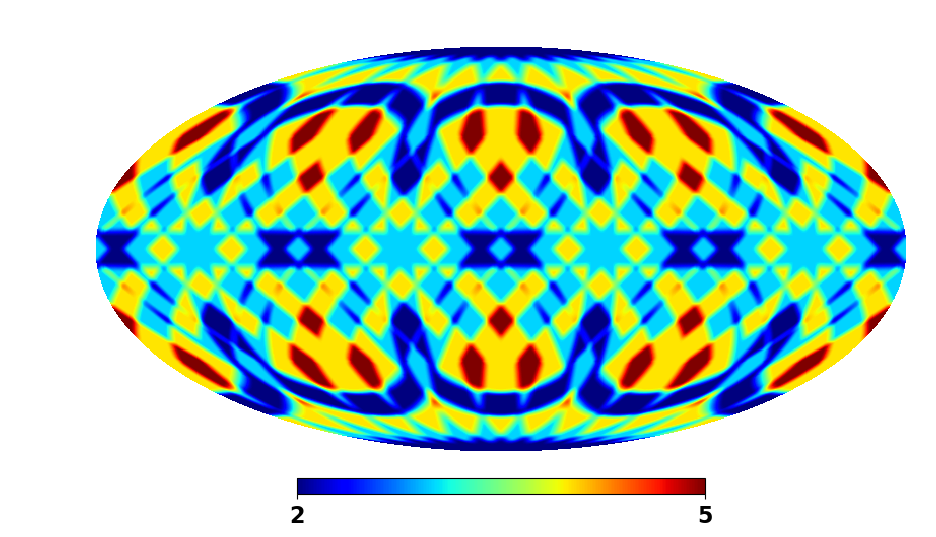}
    \end{tabular}
  \end{center}
  \caption{\label{fig:MaskCover}
           Partition of unity for the rotation-based reconstruction.
           The weights smoothly decaying toward the border are presented
           in the top left and are copied to each HEALPix face in the top right.
           In the bottom left figure, resampling was first performed
           using a rotation and bilinear interpolation,
           and the image shows the weights that would be applied
           in the original reference coordinates.
           The resulting covering of the sphere using 5 rotations
           is illustrated in the last part of the figure.}
\end{figure}

The reconstruction is performed using a partition of unity on the sphere
(see Fig.~\ref{fig:MaskCover}),
which is obtained from weights that are
smoothly decaying from $1$ in a central region of the faces
to $0$ at their borders and therefore mitigating border effects.
Formally, the reconstruction reads:
\begin{equation}
  \label{eq:rot_recons}
  \mathbf{\widetilde{X}}
  = \mathbf{N}
    \sum_{\mathbf{r}}
      \sum^{12}_{f=1}
        \mathcal{R}_{-\mathbf{r}}
        (
         \mathbf{H}^T_f
           \mathbf{M}
             \mathcal{T}_\alpha
             ( \boldsymbol{\lambda}_{\alpha,\mathbf{r},f} )
        ) \, ,
\end{equation}
where $\mathcal{R}_{-\mathbf{r}}$ resamples the sphere with the inverse rotation matrix, $\mathcal{T}_{\alpha}$ is computing the inverse $\alpha$-shearlet transform,
$\mathbf{M}$ applies weights,
and the normalization matrix $\mathbf{N}$ is chosen such that
$\mathbf{N}
 \sum_{\mathbf{r},f}
   \mathcal{R}_{-\mathbf{r}}
     \left(
       \mathbf{H}^T_f \mathbf{M} \mathbf{1}
     \right)
 =\mathbf{1}$
where $\mathbf{1}$ is a vector with all entries equal to $1$.
An example of the weights and normalization maps used to construct this
partition of unity are illustrated in Fig.~\ref{fig:MaskCover}.

Note that since the rotations $\mathcal{R}_{\mathbf{r}}$ and
$\mathcal{R}_{-\mathbf{r}}$ are implemented using interpolation,
it is not true \emph{exactly} that
$\mathcal{R}_{-\mathbf{r}} \mathcal{R}_{\mathbf{r}} \mathbf{X} = \mathbf{X}$.
Therefore, even if the coefficients $\boldsymbol{\lambda}_{\alpha,\mathbf{r},f}$
are obtained through Eq.~\eqref{eq:rot_decomp},
the reconstruction in Eq.~\eqref{eq:rot_recons}
will only satisfy $\widetilde{\mathbf{X}} \approx \mathbf{X}$, not
$\widetilde{\mathbf{X}} = \mathbf{X}$.
However, the error introduced by the inexact inverse rotation is
often negligible, at least for sufficiently smooth signals;
see Section~\ref{sub:RotationOrPatchwork} for a further comment on this.

\subsubsection{The ``patchwork'' approach}
\label{sub:Patchwork}
%
%

\begin{figure}[h]\hfill
  \begin{minipage}[b]{144pt}
  \raisebox{16pt}{
  \includegraphics[width=144pt]{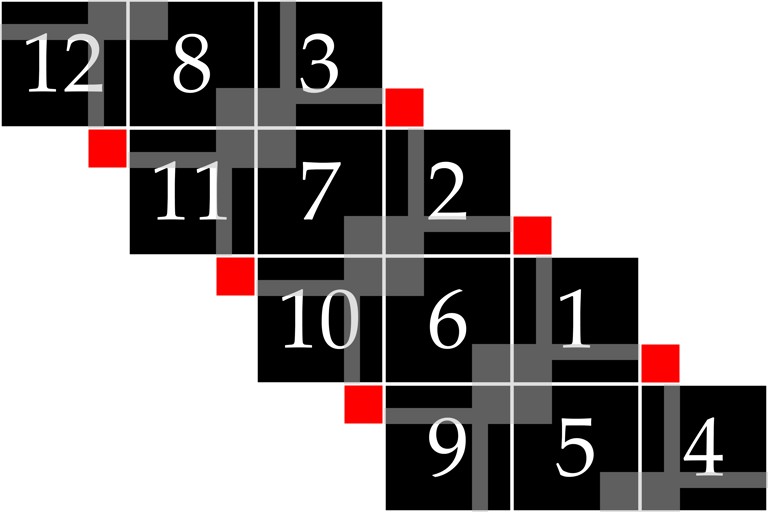}%
  }
  \end{minipage}\hfill
  \begin{minipage}[b]{86pt}
  \includegraphics[width=42pt]{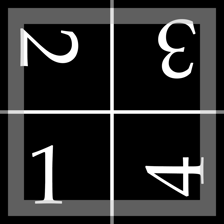}\hfill
  \includegraphics[width=42pt]{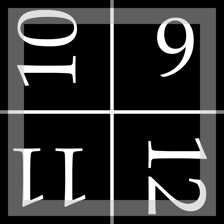}\vspace{1pt}\\
  \includegraphics[width=42pt]{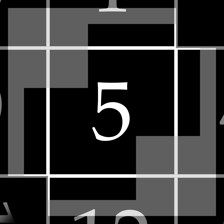}\hfill
  \includegraphics[width=42pt]{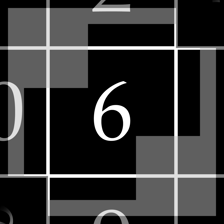}\vspace{1pt}\\
  \includegraphics[width=42pt]{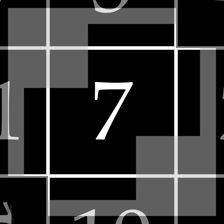}\hfill
  \includegraphics[width=42pt]{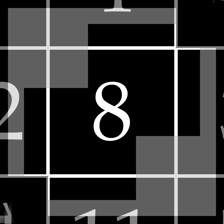}
  \end{minipage}\hfill
  \caption{Left: The twelve squares corresponding to the faces
           of the HEALPix framework (see Fig.~\ref{fig:healpix_grid})
           arranged as a net in the plane.
           The areas that are covered by multiple of the extended faces---%
           the \emph{transition zones}---are displayed in gray.
           The areas where pixels are ``missing'' are displayed in red.
           Right: The six extended faces produced by the patchwork procedure.
           The two polar faces form the top row,
           followed by the four equatorial faces below.
           The shaded area around the transition zone of each composite face
           indicates the \emph{margin}, which is later discarded.}
  \label{fig:numbers_net_transitions}
\end{figure}

The ``patchwork'' approach is another strategy to eliminate artefacts
that arise if one naively uses the disjoint HEALPix faces.
Contrary to the rotation-based technique, where an interpolation
is performed during the resampling,
the patchwork approach is based on extending the HEALPix faces
using parts of the surrounding faces so as to avoid interpolation.
Similar to the rotation-based approach, the six resulting extended faces
(see Fig.~\ref{fig:numbers_net_transitions}) form
a \emph{redundant} covering of the sphere, which is beneficial for avoiding
boundary artefacts.
Once these six extended faces are computed, the $\alpha$-shearlet transform
and all further processing are performed on these faces.
Of course, for the reconstruction, the last step consists in combining
the redundant faces to get back a proper HEALPix map.

Formally, the decomposition can be described as follows:
\begin{equation}
  \label{eq:patch_decomp}
  \boldsymbol{\lambda}_{\alpha,f}
  = \mathcal{S}_\alpha \left(\mathcal{P}_f \left(\mathbf{X}\right)\right),
\end{equation}
where $\mathcal{P}_{f}$ is now the operator that extracts the extended face $f$
from the HEALPix map $\mathbf{X}$.
Similarly, the reconstruction reads:
\begin{equation}
  \label{eq:patch_recons}
  \mathbf{\widetilde{X}}
  = \mathcal{M}
    \left[
      \left(
        \mathcal{T}_\alpha
        \left( \boldsymbol{\lambda}_{\alpha,f} \right)
      \right)_{f = 1,\dots,6}
    \right] \, ,
\end{equation}
where $\mathcal{M}$ is the operator that reconstructs a HEALPix
map from data on the six extended faces.

\begin{figure}[h]
  \begin{center}
    \includegraphics[width=100pt]{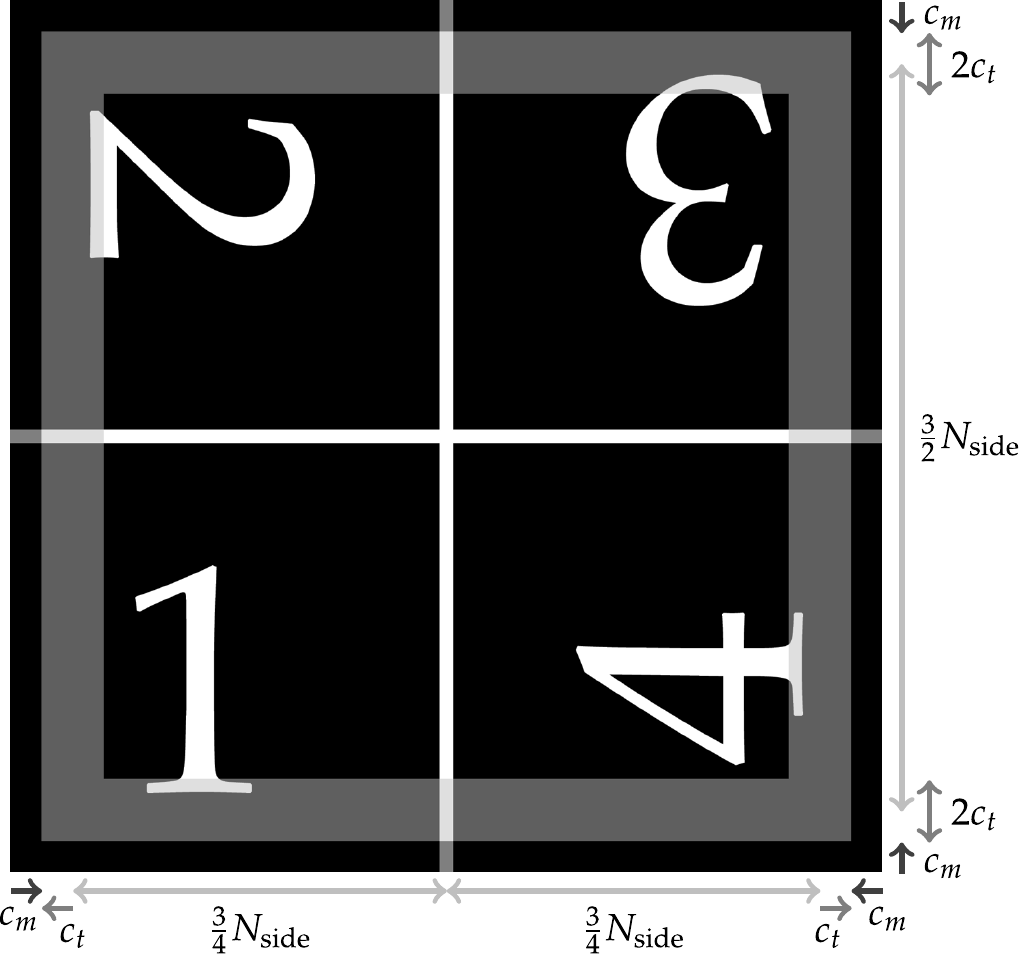}
    \qquad
    \includegraphics[width=100pt]{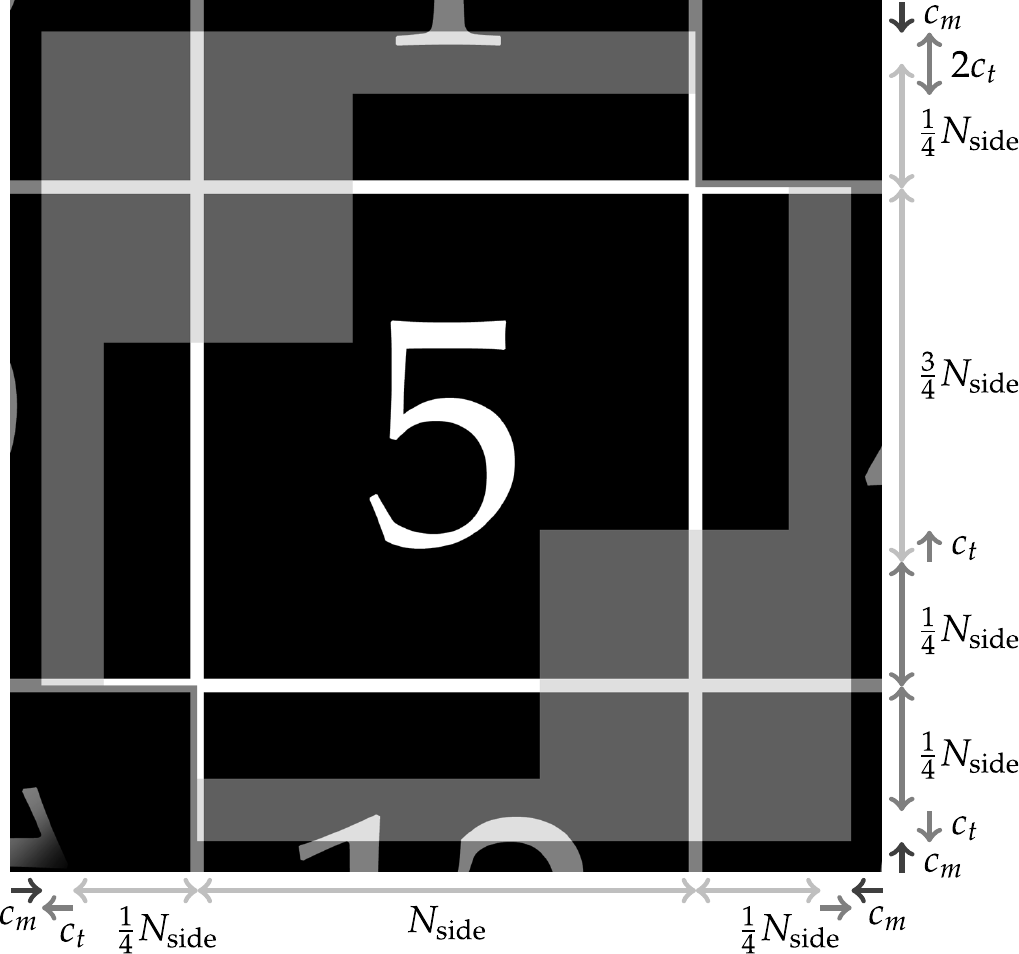}
  \end{center}
  \caption{Detailed view of two of the six extended faces.
           The dark outer boundary with width $c_m$ is the \emph{margin}
           that is discarded after the reconstruction step, and the two dark squares in the corners of the equatorial face on the right are treated likewise.
           The remaining part of the extended faces has a gray outer boundary
           of width $2 c_t$. In conjunction with the gray squares 
           in the corners of the equatorial face, this boundary forms the \emph{transition zone}
           that contains the values shared
           with the neighboring extended faces.}
  \label{fig:extended_faces_detail}
\end{figure}

The rest of this section explains how precisely the extended faces
are obtained from the original HEALPix faces, and conversely how a HEALPix
map can be obtained from data on these six extended faces.
For an accompanying visual explanation of the procedure, the reader
should consult Figures \ref{fig:healpix_grid}, \ref{fig:numbers_net_transitions},
and \ref{fig:extended_faces_detail}.

Each of the six extended faces consists of an inner square
with HEALPix pixels that are unique to this extended face,
and a border zone with HEALPix pixels that appear in several of the extended faces.
The border itself is again subdivided in an outer \emph{margin}
that is disregarded after the reconstruction step
so that the artefacts at the boundary are cut off (not mapped to the sphere),
and an inner part that forms a \emph{transition zone},
where the values of neighboring faces are blended together,
to prevent visible discontinuities between them.

Instead of extending all \emph{twelve} original faces,
we combine them to six bigger composite faces and extend those.
This reduces the number of additional pixels that have to be processed
(when using a border of the same size),
at the cost of increased memory requirements.
The first two composite faces cover the bulk of
the north and south polar regions, and particularly the poles itself.
Since the four faces of each polar region meet at the poles,
we can arrange those four faces to form a square around the pole.
It only remains to clip this area to the requested size. 
Although there is much freedom to set the extent
of the individual composite faces, we prefer all squares to be of equal size,
so that they can be processed without distinction.
The remaining four composite faces are obtained
by expanding the equatorial faces.
An expansion of the equatorial faces by $\frac{N_\text{side}}{4}$
in each direction results in areas of width $\frac{3N_\text{side}}{2}$,
that each contain a fourth of every surrounding polar face.
By removing those parts from the polar areas, constructed earlier,
those are truncated to the same width (see Fig.~\ref{fig:extended_faces_detail}).
Thus, we get six areas of equal size that cover the sphere.
Chosen this way, there is still no overlap between the polar
and equatorial composite faces; therefore we extend each face
further by half the requested width of the transition zone. We chose an extension of width $\frac{N_\text{side}}{16}$ (that is $c_{t}$ in Fig.~\ref{fig:extended_faces_detail}).
Since each face enters their neighbors territory by that amount,
this results in a transition zone of width $\frac{N_\text{side}}{8}$
between each face.
Additionally each face is extended by a margin (that is $c_{m}$ in Fig.~\ref{fig:extended_faces_detail}) to avoid border artefacts. Here, a margin of width $\frac{N_\text{side}}{16}$ was chosen.

However, to extend the equatorial faces, we have to address the problem
that there are eight vertexes where two faces of a polar region
meet a face of the equatorial region
(located on the circles of latitude $\theta = \cos^{-1}(\pm 2/3)$,
depicted in Fig.~\ref{fig:healpix_grid}).
By arranging the twelve faces as a net in the plane---as illustrated in
Fig.~\ref{fig:numbers_net_transitions}---it becomes
clear that there are gaps between the polar faces, where no values exist;
these areas are marked in red in Fig.~\ref{fig:numbers_net_transitions}.
We need to fill those gaps in order to obtain rectangular extended faces,
to which we can apply the $\alpha$-shearlet transform.
In the end, these parts will be cut away and disregarded
like the outer margin of the extension,
so the filled in values will not actually be used for the reconstruction.
Nevertheless, we need to be careful, since otherwise we might introduce
additional artefacts like the ones at the boundary.

\begin{figure}[h]\centering\hfill
  \includegraphics[width=48pt]{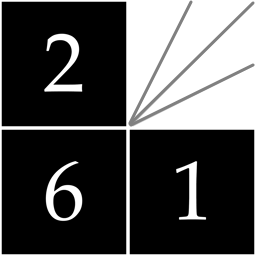}\hfill
  \includegraphics[width=48pt]{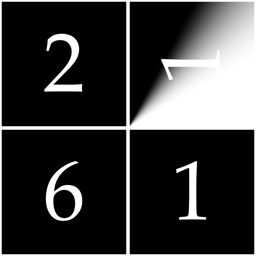}\hfill
  \includegraphics[width=48pt]{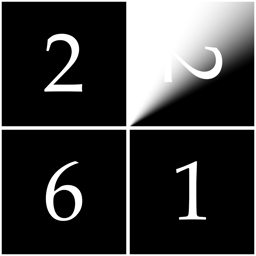}\hfill
  \includegraphics[width=48pt]{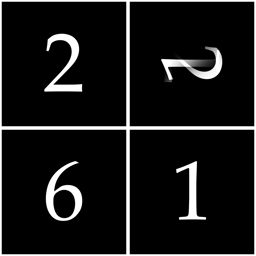}\hfill
  \caption{The ``missing'' square between faces 1 and 2
           is divided into four triangles of equal size,
           separated by the lines $2x=y$, $x=y$, and $x=2y$, as seen on the left.
           The two images in the middle reveal how the rotated faces 1 and 2
           are separately weighted along those segments:
           The data of face 1 has full weight (black) on the outer triangle
           adjacent to face 2, and no weight (white) on the other outer triangle,
           while the data of face 2 is treated conversely.
           A smooth transition is provided
           by the weights on the triangles in between.
           The sum of the weighted faces is used to fill the gap,
           as demonstrated in the right-most illustration.}
  \label{fig:filling_gap}
\end{figure}

For the sake of simplicity, we will describe the situation at the edge 
between faces 1 and 2 (see Figures \ref{fig:healpix_grid}, \ref{fig:numbers_net_transitions}, and \ref{fig:filling_gap}), which is exemplary for all gaps:
From the perspective of face 2, the missing square is expected
to feature a rotated copy of face 1,
while conversely face 1 expects a rotated copy of face 2.
To fabricate a weighted blending of those anticipated values,
we divide the empty square, interpreted as $[0,1]^2$, along the lines $2x=y$,
$x=y$, and $x=2y$, into quarters, as demonstrated in Fig.~\ref{fig:filling_gap}.
On both outer quarters the full weight is assigned to the face
which the adjoining face expects, while the two middle quarters serve
to produce a smooth transition.
All weights are normalized in such a way that every pixel is a
convex combination of the pixels of the two faces;
that is, the weights are non-negative and their sum is one at each pixel.

With this process, we fill the vertex regions with values.
Note that we don't actually need to fill the whole square,
but only the corner needed for the expansion
(the red part in Fig.~\ref{fig:numbers_net_transitions}).
Having done this, we can piece the equatorial faces together
from the various parts of the six surrounding faces and two filler squares.
Fig.~\ref{fig:numbers_net_transitions} shows the resulting
extended faces on the right.

We have now described the operators $\mathcal{P}_f$ appearing in
Eq.~\eqref{eq:patch_decomp} which assign to a given HEALPix map $\mathbf{X}$
the six extended faces
$\mathcal{P}_1 (\mathbf{X}), \dots, \mathcal{P}_6 (\mathbf{X})$.
On these \emph{rectangular} faces, we can then apply the usual $\alpha$-shearlet
transform, and do any further processing that is desired (for instance, we
can denoise the six extended faces by thresholding the
$\alpha$-shearlet coefficients).

After the processing is done on the six extended faces,
the outer margin and filler values are disregarded
and the remnant is separated along the boundaries of the original faces.
From these pieces, the original faces are put back together.
While doing so, all pixels that were part of a transition zone are weighted,
similarly as above, as a convex combination of the pixels of the
(up to four) involved extended faces.

Since we use only the values provided by the HEALPix grid,
and instead of interpolating between pixels use convex combinations of
pixel values in the transition zones, the patchwork procedure is invertible,
with Eq.~\eqref{eq:patch_recons} describing a left inverse to the
``patchwork $\alpha$-shearlet coefficient operator'' described in
Eq.~\eqref{eq:patch_decomp}.
Thus, the patchwork-based $\alpha$-shearlets form a \emph{frame}.
We emphasize, however, that the reconstruction procedure described in
Eq.~\eqref{eq:patch_recons} is not necessarily identical to the one
induced by the \emph{canonical} dual frame
of the patchwork-based $\alpha$-shearlet frame.

\section{Experiments}
\label{sect_exp}


To evaluate $\alpha$-shearlets and dictionary learning,
we have selected two different simulated data sets on the sphere:
\begin{itemize}
  \item{Thermal dust map:} a full sky thermal dust map from the
                           Planck Sky Model (100 GHz map) \citep{PlanckFFP8},
                           obtained through the Planck Legacy Archive
                           (\url{http://pla.esac.esa.int/pla/#maps}).

  \item{Horizon full sky maps:}  a series of full sky maps from the
                                 Horizon $N$-body simulations
                                 describing the dark matter halo distribution
                                 between redshift 0 and 1 \citep{Teyssier09}
                                 (see \url{http://www.projet-horizon.fr}).
\end{itemize}

While in the former scenario, the signal is smooth
and expected to be best represented by multi-scale transforms,
in the latter the signal is more discontinuous and geometrically
composed of filamentary structures joining clusters,
with density changing with redshift.
These two simulations are therefore illustrative of different scenarios
where such adaptive transforms would be useful.

\begin{figure}[h]
  \begin{center}
    \includegraphics[width=164pt]{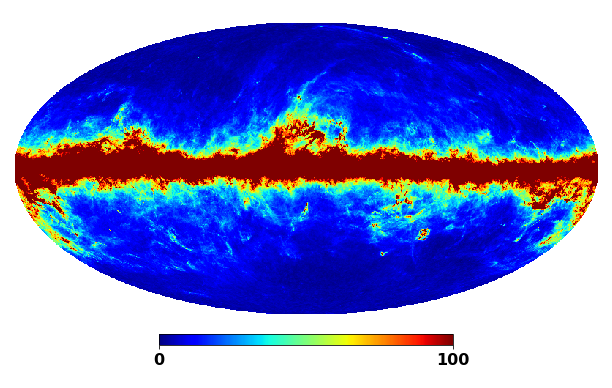}
    \includegraphics[width=164pt]{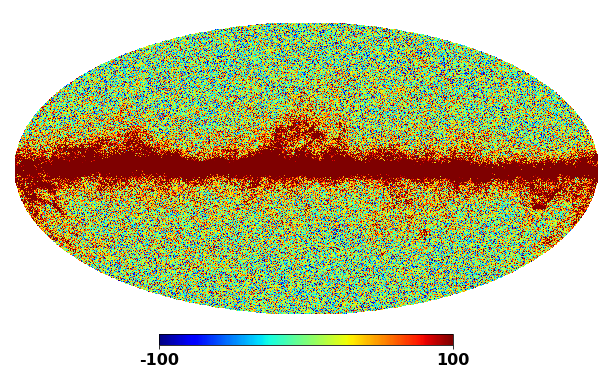}
  \end{center}
  \caption{\label{fig:TDustInputs} Thermal dust simulation map (at 100 GHZ)
                                   without (top) and with the additive
                                   white Gaussian noise added (bottom),
                                   for evaluation of the methods.
                                   The colorscale has been stretched
                                   to illustrate the challenge of recovering
                                   structures at intermediate latitude.
                                   Units in $\mu K$.}
\end{figure}

To evaluate the respective performance of DL and $\alpha$-shearlets
for denoising, we have added to the thermal dust map
an additive white Gaussian noise with standard deviation $45\mu K$,
which corresponds to the expected level of CMB at such frequency.
The resulting map can be seen in Fig.~\ref{fig:TDustInputs}.

\begin{figure}[h]
  \begin{center}
    \includegraphics[width=95pt]{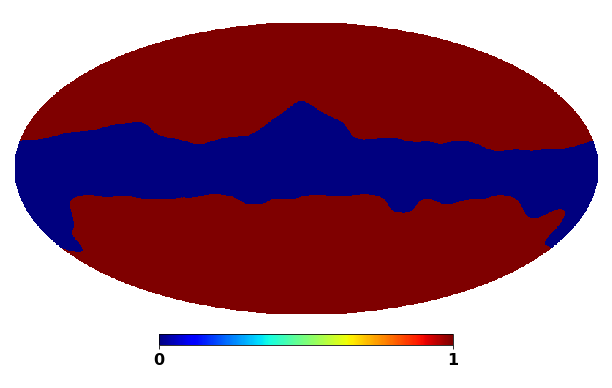}
    \includegraphics[width=95pt]{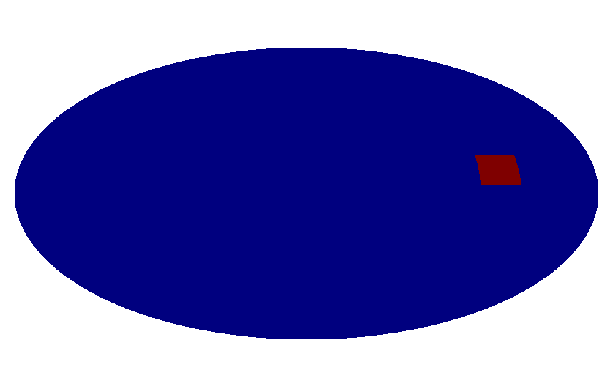}
  \end{center}
  \caption{\label{fig:GalMask} Left: galactic mask used for thermal dust
                                     quantitative evaluation,
                                     covering $70\%$ of the sky.
                               Right: region close to galactic plane
                                      where methods are inspected.}
\end{figure}

The galactic mask used for quantitative comparisons to separate regions
of high dust amplitude from regions with lower values
at higher galactic latitude is displayed in Fig.~\ref{fig:GalMask},
along with the location of a region close to the galactic plane
where the differences in between the methods could be better visualized.

\begin{figure}[h]
  \begin{center}
    \includegraphics[width=155pt]{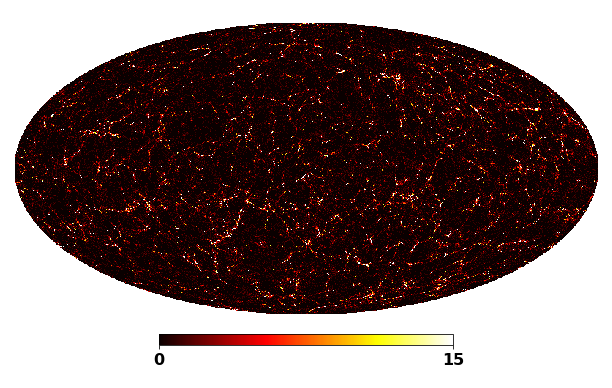}
    \includegraphics[width=155pt]{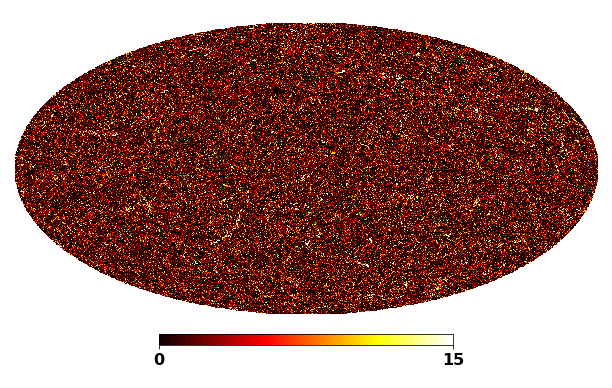}
  \end{center}
  \caption{\label{fig:DMInputs}
           Dark matter halo distribution for the first slice,
           without (top) and with the additive white Gaussian noise added (bottom),
           for evaluation of the methods.
           The colorscale has been stretched to visualize filamentary structures.
           }
\end{figure}

For the dark matter halo distribution, we select the first slice
of the data cube, and adjust the white noise level at $5$,
so that filamentary structures are of a similar amplitude as the noise,
as can be observed in Fig.~\ref{fig:DMInputs}.
This noise does not correspond to something realistic in our actual experiments,
but our goal here is only to evaluate
how different adaptive representations behave
when extracting features embedded in Gaussian noise.

In the following two subsections, we outline the precise choice of the
hyperparameters that we used, respectively, for the $\alpha$-shearlets
and for the dictionary learning based denoising.

\subsection{$\alpha$-shearlet parameters}
\label{subsec:alpha_params}
For the two $\alpha$-shearlet approaches, we used
$11$ values of $\alpha$, sampled uniformly with a density of $0.1$
ranging from $0$ to $1$.
We used $4$ scales of decomposition, using either
the rotation-based approach (Eq.~\eqref{eq:rot_decomp}),
or the patchwork approach (Eq.~\eqref{eq:patch_decomp}).
For the actual denoising, we performed a hard thresholding
of the $\alpha$-shearlet coefficients.
For this, we used different detection thresholds on different scales.
Precisely, we used a $4\sigma$ detection threshold for scale $0$
with a lower signal to noise ratio, and a detection threshold of
$3\sigma$ for the other scales; for the coarse scale,
however, we did not do any thresholding.
The reconstruction was then performed using either
Eq.~\eqref{eq:rot_recons} or \eqref{eq:patch_recons}.

For the rotation-based approach, $5$ rotations were selected
as a balance between having "more uniform" weights
and the computational burden of this approach.
The weight maps were build using a margin and transition
(smooth trigonometric variation in between 0 and 1) of size $\frac{N_\text{side}}{16}$.

For the patchwork approach, we set the size of both the utilized extension
and the margin to $\frac{N_\text{side}}{16}$, which results in increasing
the number of pixels that have to be processed by about half ($53.1\%$).
A little less than half of the added pixels are used
for the sake of redundancy, and the rest is disregarded.

\subsection{Dictionary learning parameters}
\label{subsec:DL_params}

For the thermal dust data where the information is present at several scales,
we chose the multiscale dictionary learning technique.
$3$ scales of the Starlet transform on the sphere \citep{starck:sta05_2}
were first computed from the input simulated dust map without noise.
Note that the finest wavelet scale has not been directly computed
through its spherical harmonic decomposition
to avoid artefacts for a non band-limited signal.
We followed Algorithm~\ref{Algo:DL} for the learning procedure,
with the parameters listed in Table~\ref{tab:DLDustParams}.
The patch size, the number of atoms, and the maximal sparsity
were selected experimentally by choosing values that lead
to the lowest average approximation error during the training phase.

\begin{figure}[h]
  \begin{center}
    \includegraphics[width=100pt]{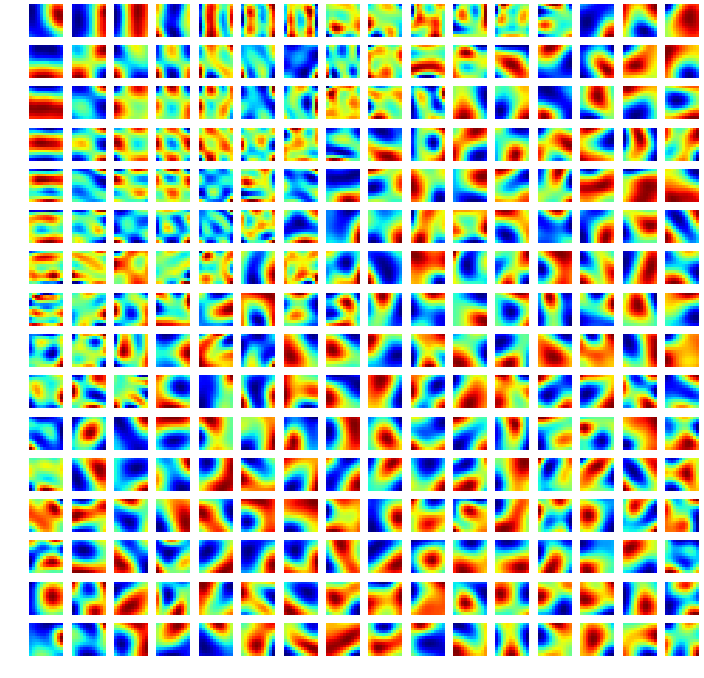}
    \includegraphics[width=100pt]{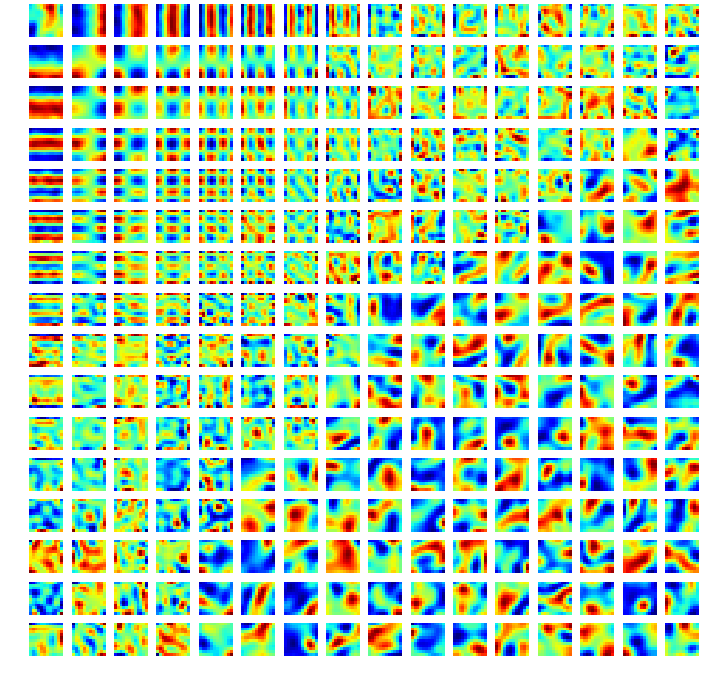}
  \end{center}
  \caption{\label{fig:TDustDico}
           Atoms learned in the multiscale dictionary learning approach.
           On the left: scale $0$, on the right: scale $1$.
           The dictionaries have departed from the original redundant DCT
           dictionary and have learned specific features related to the scale.
           Note that due to the change of the $N_{side}$ parameter with the scale,
           the actual distance between two adjacent pixels has increased,
           and the atoms for scale $1$ are indeed
           smoother than those for scale $0$.}
\end{figure}

An example of a dictionary learned for this adaptive multiscale representation
of thermal dust is shown in Fig.~\ref{fig:TDustDico}.
The dictionaries have captured at various scales both directional
and more isotropic structures.

\begin{figure}[h]
  \begin{center}
    \includegraphics[width=130pt]{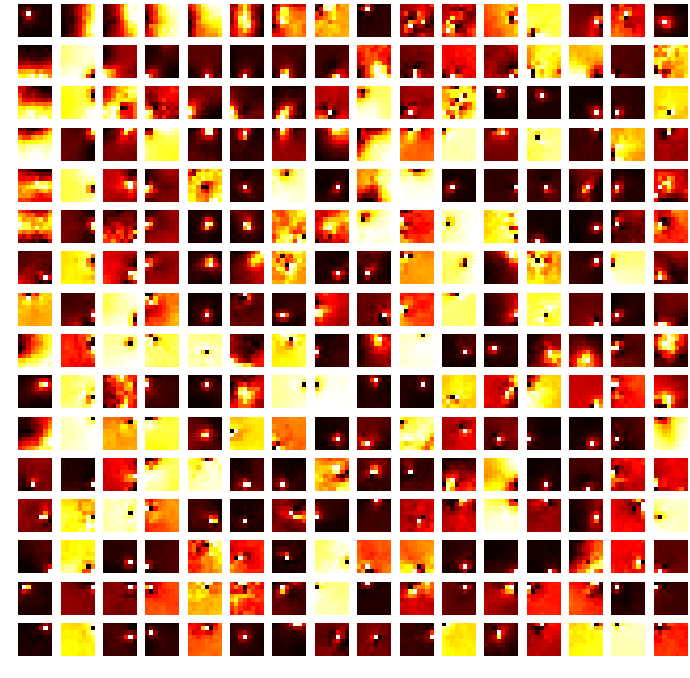}
  \end{center}
  \caption{\label{fig:DMDico}
           Atoms learned in the dictionary learning approach, applied to
           the dark matter halo distribution data.
           The dictionary elements are composed of point-like structures and edges.}
\end{figure}

In the second scenario, because information is localized in space,
the dictionary was learned directly on patches extracted from the first slice
describing the dark matter halo distribution,
from a training set of $200,000$ patches of size $8\times 8$.
As in the previous experiment,
a stopping criterion was set for the approximation error
(which should be less than the targeted level of noise),
and a maximal sparsity of $7$ was set for OMP.
K-SVD was then run for $100$ iterations.
The learned dictionary is presented in Fig.~\ref{fig:DMDico}.
Note that the atoms are essentially containing high frequency information
in this case, in contrast to the previously learned distribution
on thermal dust.

Once these dictionary are learned, the sparse decomposition step
with this representation is used for denoising.
The same parameters as above were used for the sparse coding,
except for the targeted approximation error which was set to a value
that would not be exceeded by a patch of pure noise
with a probability of $0.9545$.

\section{Results}

\label{sec:results}

\subsection{Denoising Experiments}

We tested our adaptive approaches to denoise the data in the two denoising scenario presented
in the previous section, using the parameters described in sections~\ref{subsec:alpha_params} 
and \ref{subsec:DL_params}. 

\begin{figure}[h]
  \begin{center}
    \includegraphics[width=200pt]{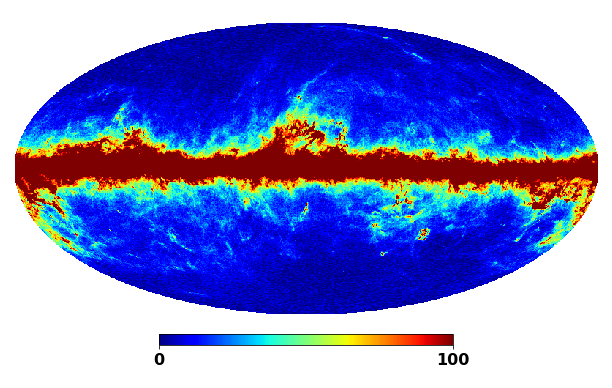}
    \includegraphics[width=200pt]{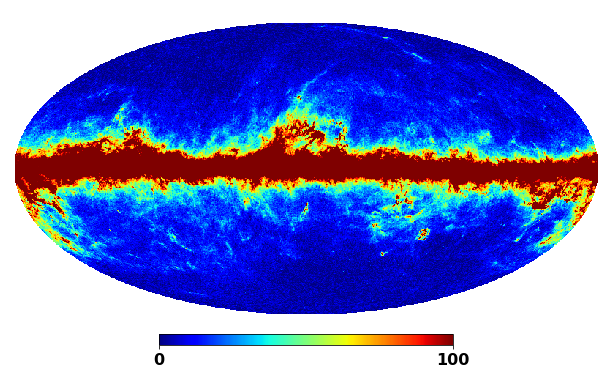}
    \includegraphics[width=200pt]{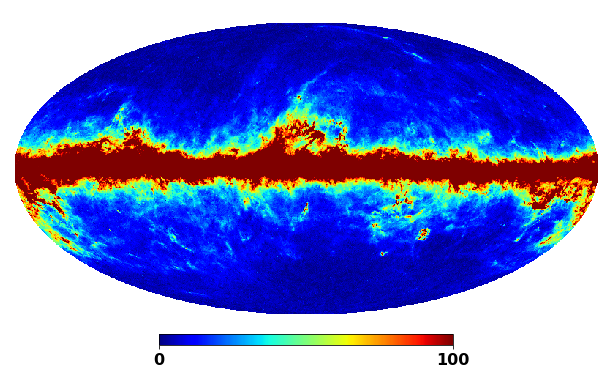}
  \end{center}
  \caption{\label{fig:TDustMaps}
           Denoised Thermal Dust Maps for all three approaches.
           Top and middle: $\alpha$-shearlet denoising with rotation-based (top)
           or  patchwork (middle) approach, both for $\alpha=0.6$;
           bottom: representation learned with dictionary learning.
           Units in $\mu K$.}
\end{figure}

\begin{figure}[h]
  \begin{center}
    \includegraphics[width=200pt]{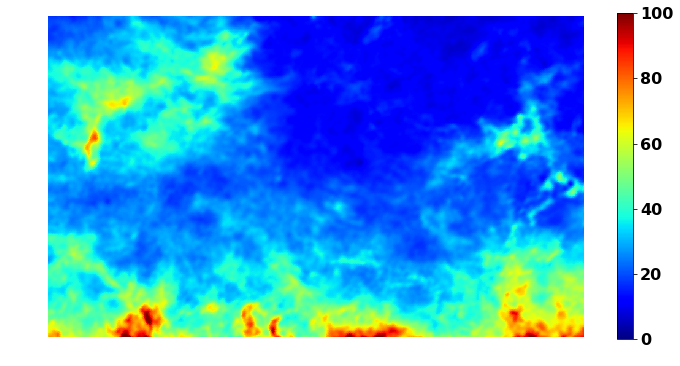}
    \includegraphics[width=200pt]{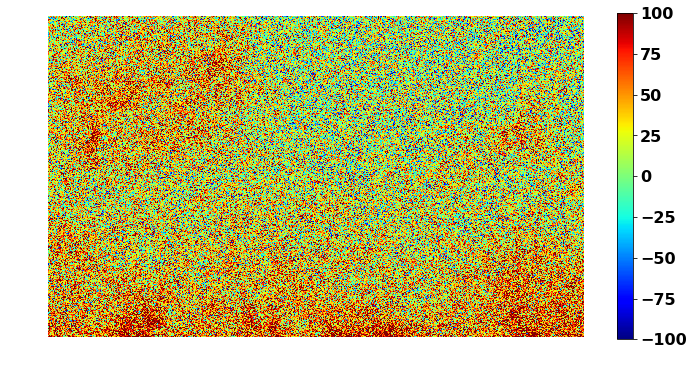}
    \includegraphics[width=200pt]{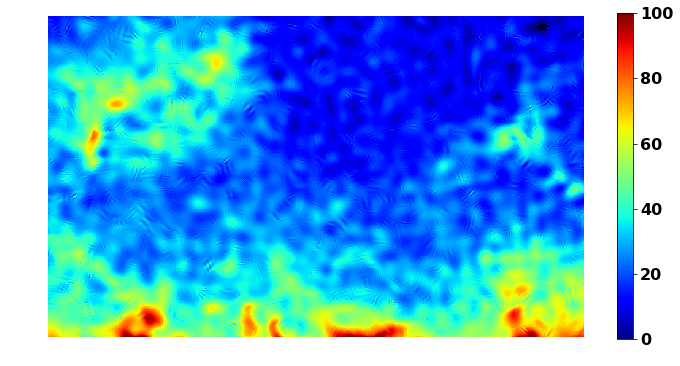}
    \includegraphics[width=200pt]{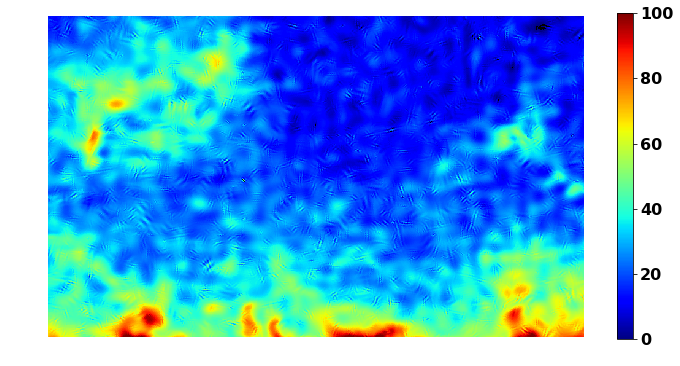}
    \includegraphics[width=200pt]{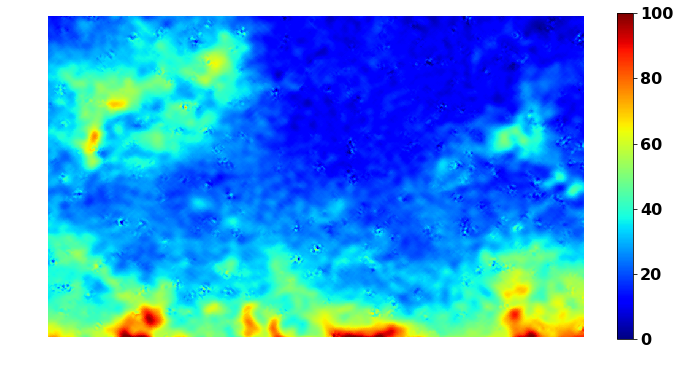}
  \end{center}
  \caption{\label{fig:TDustPyxis}
           Zoom on a region close to the galactic plane
           to visualize the respective denoising performance of the methods.
           From top to bottom: input map,
           noisy map (with own colorscale),
           rotation-based approach with $\alpha=0.6$,
           patchwork approach with $\alpha=0.6$,
           sparse representation learned from data. 
           All units are in $\mu K$.
           }
\end{figure}

\begin{figure}[h]
  \begin{center}
    \includegraphics[width=200pt]{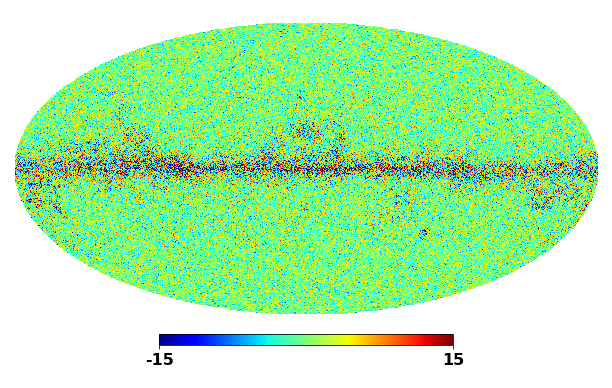}
    \includegraphics[width=200pt]{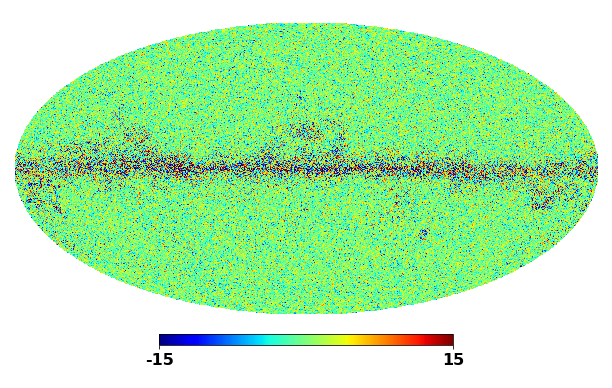}
    \includegraphics[width=200pt]{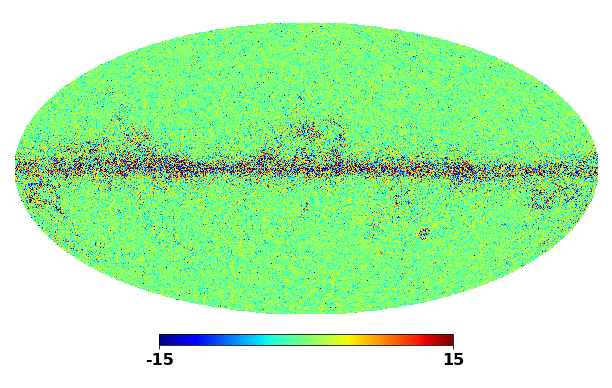}
  \end{center}
  \caption{\label{fig:TDustResiduals}
           Residuals for the maps displayed in Fig.~\ref{fig:TDustMaps}.
           Units in $\mu K$.}
\end{figure}

For the thermal dust simulation, the full sky denoised maps using the three approaches are displayed
in Fig.~\ref{fig:TDustMaps}, with a zoom to a region close to the galactic plane
in Fig.~\ref{fig:TDustPyxis} to visually inspect the differences between methods. 
Residuals on the full sphere are also shown in Fig.~\ref{fig:TDustResiduals}, 
and the performance of each approach are quantitatively evaluated in Table \ref{tab:ResDust}
in the full sky as well as in regions defined by the galactic mask.

\begin{table*}[htb]
  \caption{\label{tab:ResDust}
           Statistics on the recovery of spherical thermal dust maps
           with the proposed approaches.
           Bias, root mean square error (RMSE)
           and mean absolute deviation (MAD) are presented,
           for the overall map (All),
           the region not in the mask (Out)
           and the galactic region (Gal.) defined by the mask of Fig.~\ref{fig:GalMask}.
           The best results are in bold,
           the best results among $\alpha$-shearlets are underlined.
           Units in $\mu K$.}
  \begin{center}
  \begin{tabular}{|l|l|l|l|l|l|l|l|l|l|l|}
  \hline
  \multicolumn{2}{|c|}{Method} & \multicolumn{3}{|c|}{Bias} & \multicolumn{3}{|c|}{RMSE} & \multicolumn{3}{|c|}{MAD} \rule[-1ex]{0pt}{3.5ex}\\
  \multicolumn{2}{|c|}{     } & All & Out & Gal.    &  All & Out &  Gal.   &  All & Out & Gal. \rule[-1ex]{0pt}{3.5ex}\\
  \hline
  \rule[-1ex]{0pt}{3.5ex}\multirow{11}{*}{Rotation} &$\alpha=0  $ & 0.008 & 0.005 & 0.016 & 4.266 & 3.028 & 6.270 & 3.020 & 2.392 & 4.490\\
  \rule[-1ex]{0pt}{3.5ex}                           &$\alpha=0.1$ & 0.008 & 0.005 & 0.016 & 4.264 & 3.025 & 6.268 & 3.018 & 2.389 & 4.488\\
  \rule[-1ex]{0pt}{3.5ex}                           &$\alpha=0.2$ & 0.008 & 0.005 & 0.016 & 4.261 & 3.022 & 6.264 & 3.016 & 2.387 & 4.485\\
  \rule[-1ex]{0pt}{3.5ex}                           &$\alpha=0.3$ & 0.008 & 0.005 & 0.016 & 4.256 & 3.019 & 6.257 & 3.012 & 2.384 & 4.480\\
  \rule[-1ex]{0pt}{3.5ex}                           &$\alpha=0.4$ & 0.008 & 0.005 & 0.016 & 4.256 & 3.021 & 6.255 & 3.012 & 2.384 & 4.480\\
  \rule[-1ex]{0pt}{3.5ex}                           &$\alpha=0.5$ & 0.008 & 0.005 & 0.016 & 4.258 & 3.024 & 6.257 & 3.012 & 2.384 & 4.481\\
  \rule[-1ex]{0pt}{3.5ex}                           &$\alpha=0.6$ & 0.008 & 0.005 & 0.016 & \underline{4.252} & \underline{3.017} & \underline{\textbf{6.252}} & \underline{3.008} & \underline{2.380} & \underline{\textbf{4.477}}\\
  \rule[-1ex]{0pt}{3.5ex}                           &$\alpha=0.7$ & 0.008 & 0.005 & 0.016 & 4.256 & 3.020 & 6.256 & 3.010 & 2.381 & 4.480\\
  \rule[-1ex]{0pt}{3.5ex}                           &$\alpha=0.8$ & 0.008 & 0.005 & 0.016 & 4.257 & 3.019 & 6.261 & 3.010 & 2.380 & 4.483\\
  \rule[-1ex]{0pt}{3.5ex}                           &$\alpha=0.9$ & 0.008 & 0.005 & 0.016 & 4.260 & 3.019 & 6.266 & 3.011 & 2.380 & 4.486\\
  \rule[-1ex]{0pt}{3.5ex}                           &$\alpha=1  $ & 0.008 & 0.005 & 0.016 & 4.267 & 3.027 & 6.273 & 3.012 & 2.380 & 4.489\\
  \hline
  \rule[-1ex]{0pt}{3.5ex}\multirow{11}{*}{Patchwork}&$\alpha=0  $ & 0.008 & 0.006 & 0.014 & 4.507 & 3.383 & 6.409 & 3.252 & 2.657 & 4.643\\
  \rule[-1ex]{0pt}{3.5ex}                           &$\alpha=0.1$ & 0.008 & 0.006 & 0.014 & 4.502 & 3.376 & 6.404 & 3.246 & 2.650 & 4.638\\
  \rule[-1ex]{0pt}{3.5ex}                           &$\alpha=0.2$ & 0.008 & 0.006 & 0.014 & 4.499 & 3.375 & 6.398 & 3.243 & 2.648 & 4.634\\
  \rule[-1ex]{0pt}{3.5ex}                           &$\alpha=0.3$ & 0.008 & 0.006 & 0.014 & 4.488 & \underline{3.364} & 6.386 & 3.231 & 2.636 & 4.642\\
  \rule[-1ex]{0pt}{3.5ex}                           &$\alpha=0.4$ & 0.008 & 0.006 & 0.014 & 4.492 & 3.373 & 6.385 & 3.235 & 2.641 & 4.624\\
  \rule[-1ex]{0pt}{3.5ex}                           &$\alpha=0.5$ & 0.008 & 0.006 & 0.014 & 4.497 & 3.379 & 6.388 & 3.232 & 2.637 & 4.623\\
  \rule[-1ex]{0pt}{3.5ex}                           &$\alpha=0.6$ & 0.008 & 0.006 & 0.014 & \underline{4.485} & 3.366 & \underline{6.377} & \underline{3.223} & \underline{2.628} & \underline{4.615}\\
  \rule[-1ex]{0pt}{3.5ex}                           &$\alpha=0.7$ & 0.008 & 0.006 & 0.014 & 4.497 & 3.382 & 6.385 & 3.230 & 2.635 & 4.621\\
  \rule[-1ex]{0pt}{3.5ex}                           &$\alpha=0.8$ & 0.008 & 0.006 & 0.014 & 4.502 & 3.388 & 6.390 & 3.234 & 2.639 & 4.626\\
  \rule[-1ex]{0pt}{3.5ex}                           &$\alpha=0.9$ & 0.008 & 0.006 & 0.014 & 4.509 & 3.395 & 6.398 & 3.239 & 2.644 & 4.632\\
  \rule[-1ex]{0pt}{3.5ex}                           &$\alpha=1  $ & 0.008 & 0.006 & 0.014 & 4.527 & 3.416 & 6.413 & 3.233 & 2.634 & 4.633\\
  \hline
  \multicolumn{2}{|c|}{Dict. Learn.}                              & 0.008 & 0.006 & 0.014 & \textbf{4.034} & \textbf{2.343} & 6.440 & \textbf{2.570} & \textbf{1.750} & 4.487\rule[-1ex]{0pt}{3.5ex}\\
  \hline
  \end{tabular}
  \end{center}
\end{table*}

\begin{figure}[h]
  \begin{center}
    \includegraphics[width=200pt]{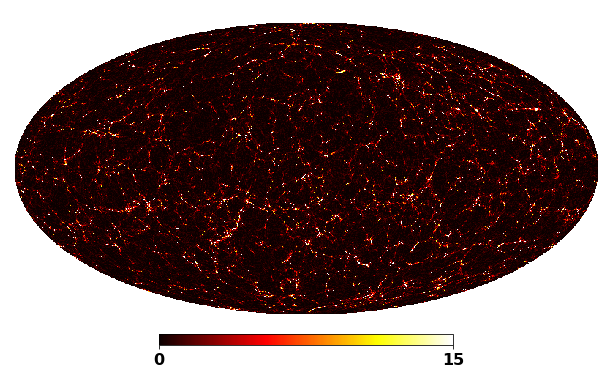}
    \includegraphics[width=200pt]{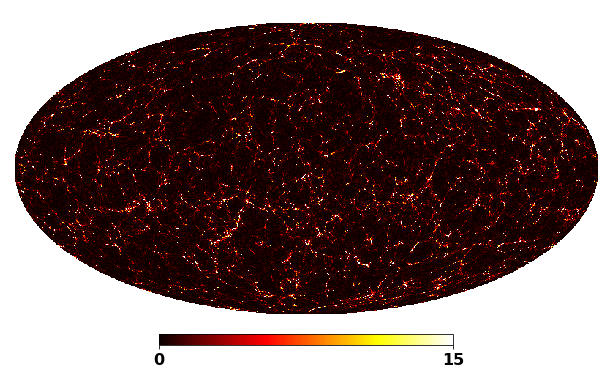}
    \includegraphics[width=200pt]{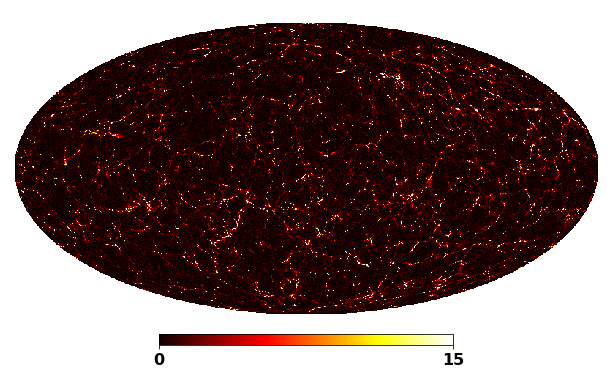}
  \end{center}
  \caption{\label{fig:DMMaps}
           Denoised dark matter maps for all three approaches.
           Top and middle:
           $\alpha$-shearlet denoising with rotation-based (top)
           or patchwork (middle) approach, both with $\alpha=1$;
           bottom: representation learned with dictionary learning.}
\end{figure}

\begin{figure}[h]
  \begin{center}
    \includegraphics[width=200pt]{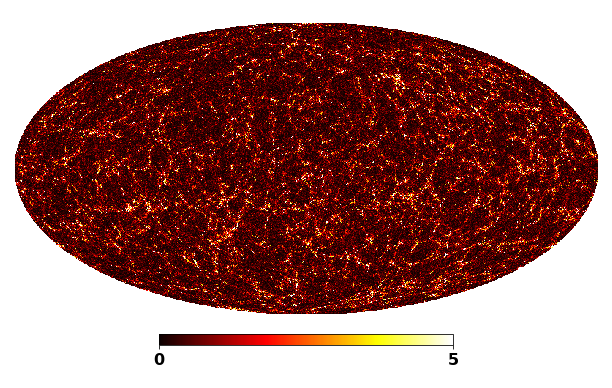}
    \includegraphics[width=200pt]{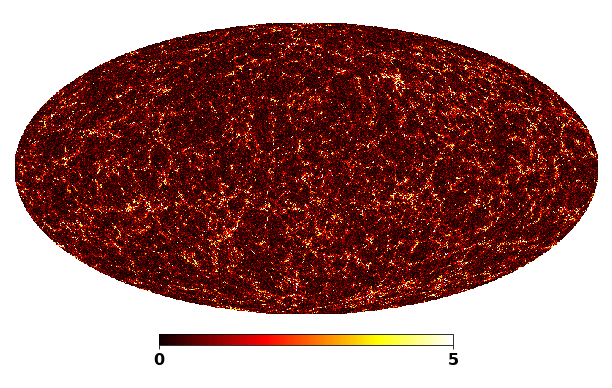}
    \includegraphics[width=200pt]{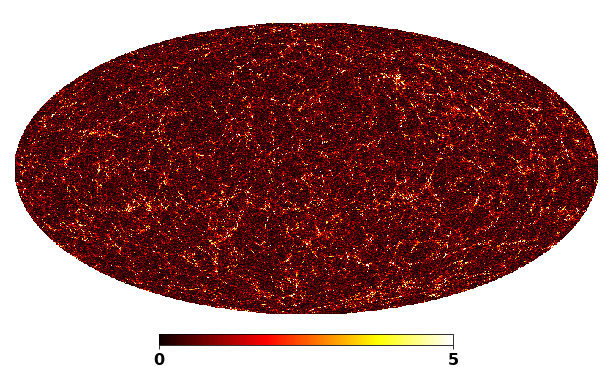}
  \end{center}
  \caption{\label{fig:DMRes}
           Amplitude of the residuals for all three approaches, for the dark matter map scenario.
           Top and middle:
           $\alpha$-shearlet denoising with rotation-based (top)
           or patchwork (middle) approach, both with $\alpha=1$;
           bottom: representation learned with dictionary learning.}
\end{figure}

\begin{figure}[h]
  \begin{center}
  \begin{tabular}{ll}
    \includegraphics[width=77pt]{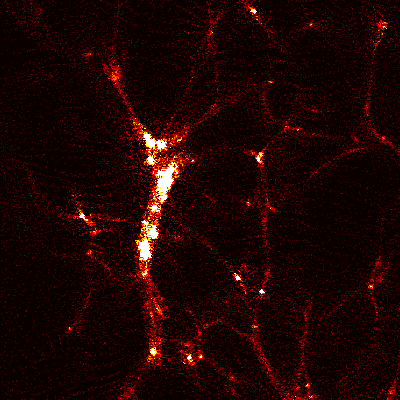}
    \includegraphics[width=77pt]{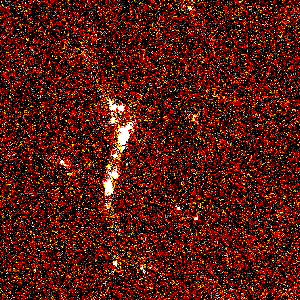}
    \includegraphics[width=77pt]{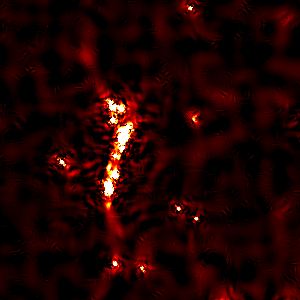}&\\
    \includegraphics[width=77pt]{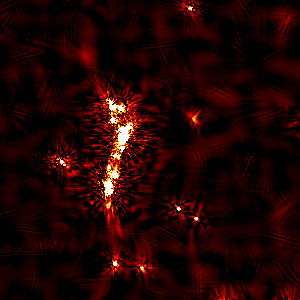}
    \includegraphics[width=77pt]{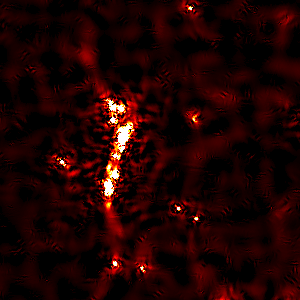}
    \includegraphics[width=77pt]{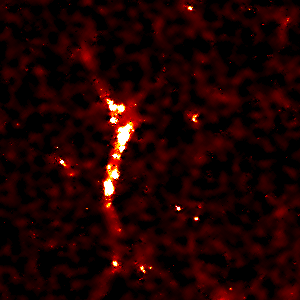}&    \includegraphics[height=77pt]{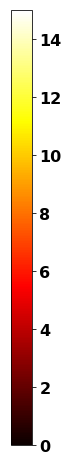}
  \end{tabular}

  \end{center}
  \caption{\label{fig:DMMapsZoom}
           Dark Matter Map amplitudes
           for all three approaches in a zoomed region.
           From top to bottom and left to right:
           original map,
           noisy map,
           rotation-based approach with $\alpha=1$,
           patchwork approach with $\alpha=0$,
           patchwork approach with $\alpha=1$,
           representation learned from data.
           }
\end{figure}

Similarly, for the dark matter halo distribution, the  full sky denoised maps are displayed
in Fig.~\ref{fig:DMMaps} and the residuals are presented in Fig.~\ref{fig:DMRes}. 
To better inspect the recovery of the filamentary structures as well as the core regions, 
a zoom in was also performed for this dataset in Fig.~ \ref{fig:DMMapsZoom}. 
Finally, the results were quantitatively evaluated in Table \ref{tab:ResDM}. 

To inspect the impact of the anisotropy parameter on the recovery of geometrical structures 
in the different redshift slices, we also computed for the patchwork approach the non-linear 
approximation curves which display the evolution of the RMSE as a function of given thresholds. 
This allows to give a more comprehensive view of the best $\alpha$ for different density levels thresholds. 
These non-linear approximation curves are
illustrated in linear and log scale in Figs~\ref{fig:DMNLA}
and \ref{fig:DMNLA_LOG}, respectively.

\begin{table}[htb]
  \caption{\label{tab:ResDM}
           Statistics on the recovery of dark matter halo distribution
           with the proposed approaches.
           Bias,
           root mean square error (RMSE)
           and mean absolute deviation (MAD) are presented.
           The best results for RMSE and MAD are in bold,
           the best results among $\alpha$-shearlets are underlined.}
  \begin{center}
    \begin{tabular}{|l|l|l|l|l|l|l|l|l|l|l|}
    \hline
    \multicolumn{2}{|c|}{Method} & Bias & RMSE & MAD \rule[-1ex]{0pt}{3.5ex}\\
    \hline
    \rule[-1ex]{0pt}{3.5ex}\multirow{11}{*}{Rotation} &$\alpha=0  $ & 0.0002 & 3.09 & 0.83 \\
    \rule[-1ex]{0pt}{3.5ex}                           &$\alpha=0.1$ & 0.0002 & 3.05 & 0.81 \\
    \rule[-1ex]{0pt}{3.5ex}                           &$\alpha=0.2$ & 0.0002 & 3.02 & 0.80 \\
    \rule[-1ex]{0pt}{3.5ex}                           &$\alpha=0.3$ & 0.0002 & 3.00 & 0.80 \\
    \rule[-1ex]{0pt}{3.5ex}                           &$\alpha=0.4$ & 0.0002 & 2.97 & 0.79 \\
    \rule[-1ex]{0pt}{3.5ex}                           &$\alpha=0.5$ & 0.0002 & 2.95 & 0.78 \\
    \rule[-1ex]{0pt}{3.5ex}                           &$\alpha=0.6$ & 0.0002 & 2.94 & 0.78 \\
    \rule[-1ex]{0pt}{3.5ex}                           &$\alpha=0.7$ & 0.0002 & 2.92 & 0.77 \\
    \rule[-1ex]{0pt}{3.5ex}                           &$\alpha=0.8$ & 0.0002 & 2.92 & 0.77 \\
    \rule[-1ex]{0pt}{3.5ex}                           &$\alpha=0.9$ & 0.0002 & 2.91 & 0.77 \\
    \rule[-1ex]{0pt}{3.5ex}                           &$\alpha=1  $ & 0.0002 & \underline{2.90} & \underline{0.77} \\
    \hline
    \rule[-1ex]{0pt}{3.5ex}\multirow{11}{*}{Patchwork}&$\alpha=0  $ & 0.0002 & 1.64 & 0.86 \\
    \rule[-1ex]{0pt}{3.5ex}                           &$\alpha=0.1$ & 0.0002 & 1.58 & 0.84 \\
    \rule[-1ex]{0pt}{3.5ex}                           &$\alpha=0.2$ & 0.0002 & 1.53 & 0.82 \\
    \rule[-1ex]{0pt}{3.5ex}                           &$\alpha=0.3$ & 0.0002 & 1.49 & 0.81 \\
    \rule[-1ex]{0pt}{3.5ex}                           &$\alpha=0.4$ & 0.0002 & 1.45 & 0.80 \\
    \rule[-1ex]{0pt}{3.5ex}                           &$\alpha=0.5$ & 0.0002 & 1.43 & 0.79 \\
    \rule[-1ex]{0pt}{3.5ex}                           &$\alpha=0.6$ & 0.0002 & 1.39 & 0.78 \\
    \rule[-1ex]{0pt}{3.5ex}                           &$\alpha=0.7$ & 0.0002 & 1.37 & 0.78 \\
    \rule[-1ex]{0pt}{3.5ex}                           &$\alpha=0.8$ & 0.0002 & 1.35 & 0.77 \\
    \rule[-1ex]{0pt}{3.5ex}                           &$\alpha=0.9$ & 0.0002 & \underline{1.34} & 0.77 \\
    \rule[-1ex]{0pt}{3.5ex}                           &$\alpha=1  $ & 0.0002 & 1.35 & \underline{0.77} \\
    \hline
    \multicolumn{2}{|c|}{Dict. Learn.}                              & 0.0002 & \textbf{1.32} & \textbf{0.72}\rule[-1ex]{0pt}{3.5ex}\\
    \hline
    \end{tabular}
  \end{center}
\end{table}

\subsection{Discussion}

In the following, we discuss several questions concerning the results;
in particular, we analyze the relative performance of our different approaches
to sparsifying representations on the sphere.

\subsubsection*{Block artefacts}

\begin{figure}[h]
  \begin{center}
    \includegraphics[width=77pt]{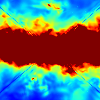}
    \includegraphics[width=77pt]{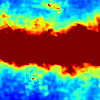}
    \includegraphics[height=77pt]{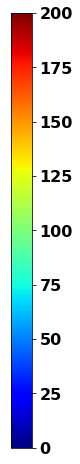}
    \\[0.5cm]
    \includegraphics[width=77pt]{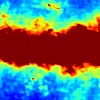}
    \includegraphics[width=77pt]{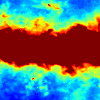}
    \includegraphics[height=77pt]{denoised_ffp8_thermaldust_nobpm_100_full_map_rnd_0_lt_DL_learn1sig_dec0p9545_10x20x30sp_nocoarse__zoom2_cart_jet_cbar}
  \end{center}
  \caption{\label{fig:TDustArtefacts}
           Cartesian projection of the denoised thermal dust maps
           centered at the intersection of 4 faces.
           From left to right and top to bottom:
           denoising each face independently using $\alpha$-shearlets with $\alpha=1$,
           restoration via the rotation-based approach,
           patchwork approach with $\alpha=1$,
           dictionary learning with patch width of 12.
           The colorscale has been stretched to visualize the artefacts
           seen as a cross-shape discontinuity
           at the boundaries of the 4 HEALPix faces in the upper left figure.
           All of our proposed approaches are free from these artefacts. Units in $\mu K$.}
\end{figure}

The first challenge in extending the representation from the Euclidean framework
to data defined on the sphere was to avoid the border effects
due to considering disjoint charts processed independently.
Fig.~\ref{fig:TDustArtefacts} illustrates that all our proposed
redundant representations, based on different overlapping charts,
are free of these block artefacts when denoising the thermal dust map. 
A similar result is obtained for denoising the dark matter maps.

\subsubsection*{Visual inspection}

Qualitatively, Figs~\ref{fig:TDustPyxis} and \ref{fig:DMMapsZoom}
illustrate the different shapes captured by 
$\alpha$-shearlets and dictionary learning atoms. In particular, for the thermal dust maps, 
the noise appears as curvelet-like structures for the former and more 
isotropic structures for the dictionary learning approach.

For the first slice of the dark matter halo distribution simulations, 
the dictionary learning approach visually seems to best recover
the structures in the data, in particular the filamentary structures and the compact cores.

\subsubsection*{Which approach is best?}

This is confirmed quantitatively in Tables \ref{tab:ResDust} and \ref{tab:ResDM}
where the dictionary learning approach outperforms overall
both $\alpha$-shearlet techniques in the denoising of thermal dust
(with a multiscale approach) and dark matter halo distribution.
For thermal dust, when looking at specific regions
(region inside or outside the galactic mask), the rotation-based approach
gives however the lowest residuals in the galactic region,
while using the learned representation gave the best results outside this region.
This could be explained by the wide diversity of amplitudes
in the galactic plane, not captured in our training set of $200,000$ patches
for the first wavelet scale, which corresponds only to $0.4\%$
of the total number of patches over the full sky.
Improving performance for dictionary learning in the galactic region
would require either to train the dictionary with a larger training set
so that it encompasses more patches from the galactic center,
or to sample more densely the galactic region
than higher galactic latitudes in this training set.

\subsubsection*{Is the rotation-based or the patchwork approach preferable?}
\label{sub:RotationOrPatchwork}

The rotation-based approach outperforms the patchwork approach
in the thermal dust denoising scenario, but conversely the patchwork approach
outperforms the rotation-based technique
in the dark matter halo distribution scenario.
The last result is due to the bilinear interpolation performed when
resampling the sphere with rotations,
which leads to severe approximation errors when the signal varies greatly
at the scale of a few pixels.

\subsubsection*{What is the best $\alpha$-value?}

Tables~\ref{tab:ResDust} and \ref{tab:ResDM} show that
for $\alpha$-shearlets in the denoising of thermal dust,
$\alpha=0.6$ (system close to the curvelets) gives the best performance,
while for the dark matter halo distribution scenario,
$\alpha=1.0$ (system close to the wavelets)
gave the best performance.

However, the second scenario displays a diversity of structures
with both high density cores and numerous less dense filaments,
with distribution changing in different slices of data corresponding to different redshifts.
It would therefore be reductive to investigate a single noise level
scenario to set a best $\alpha$ for one of this slice.

\begin{figure}[h]
  \begin{center}
    \includegraphics[width=120pt]{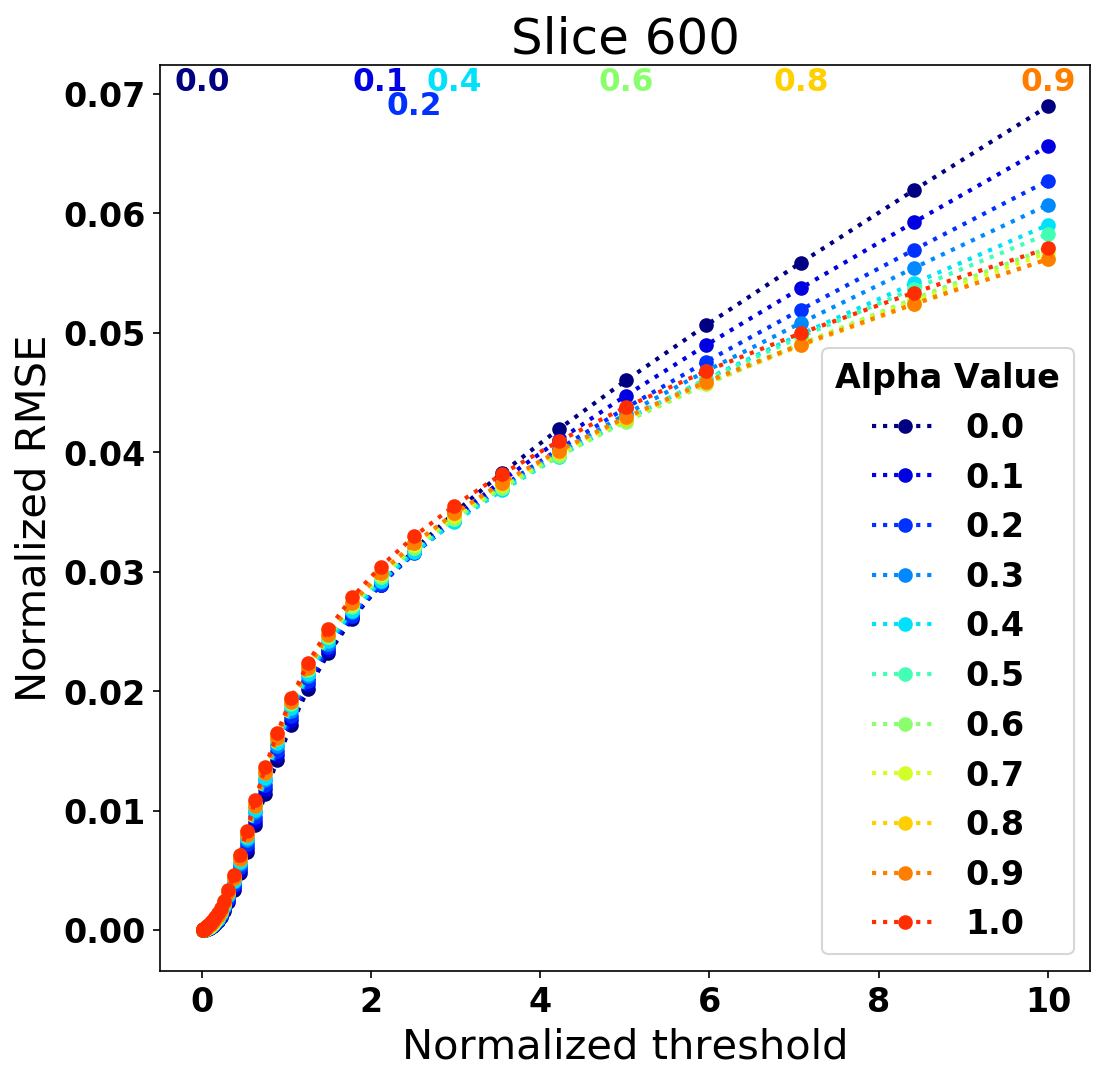}
    \includegraphics[width=120pt]{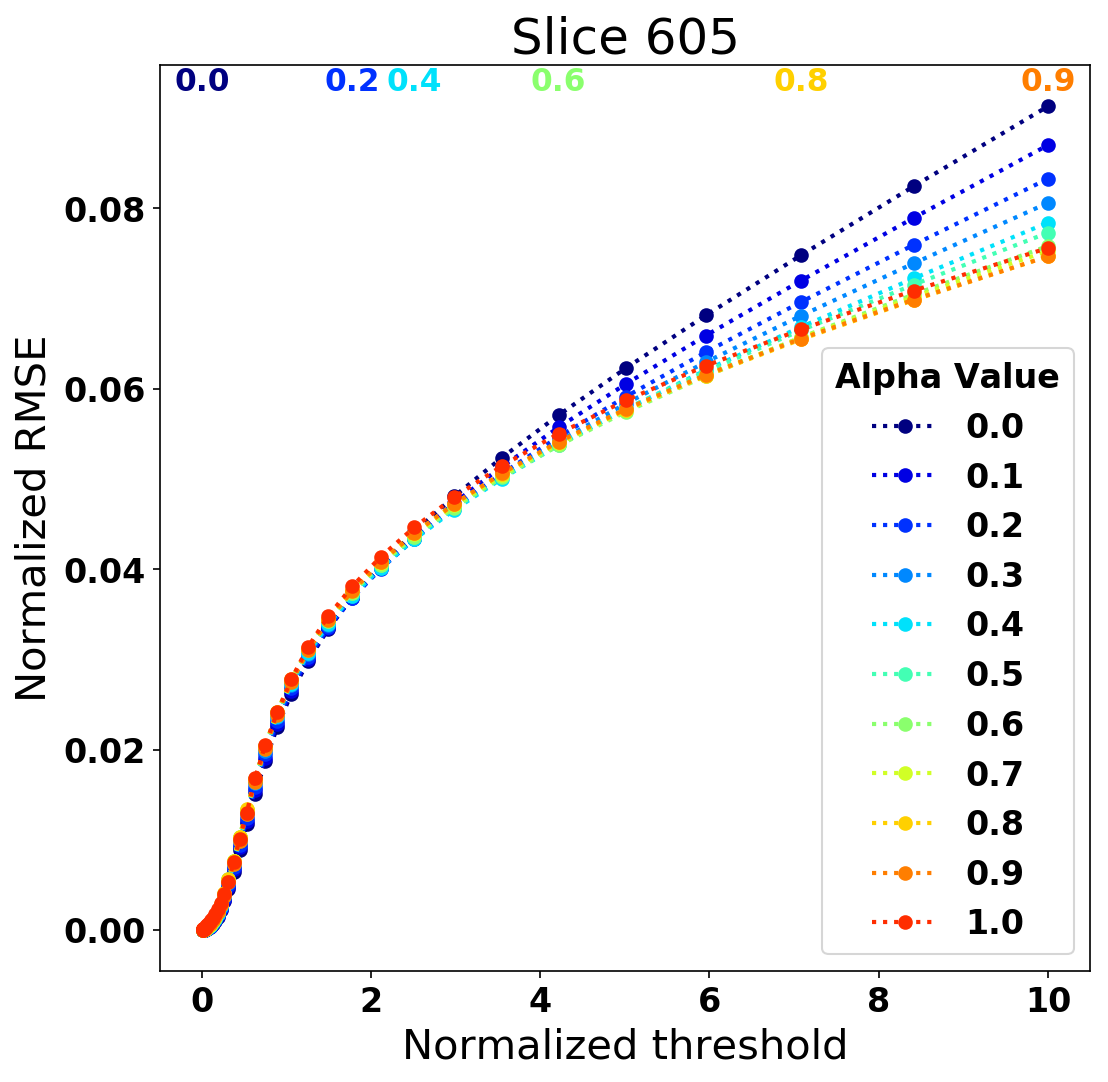}
    \includegraphics[width=120pt]{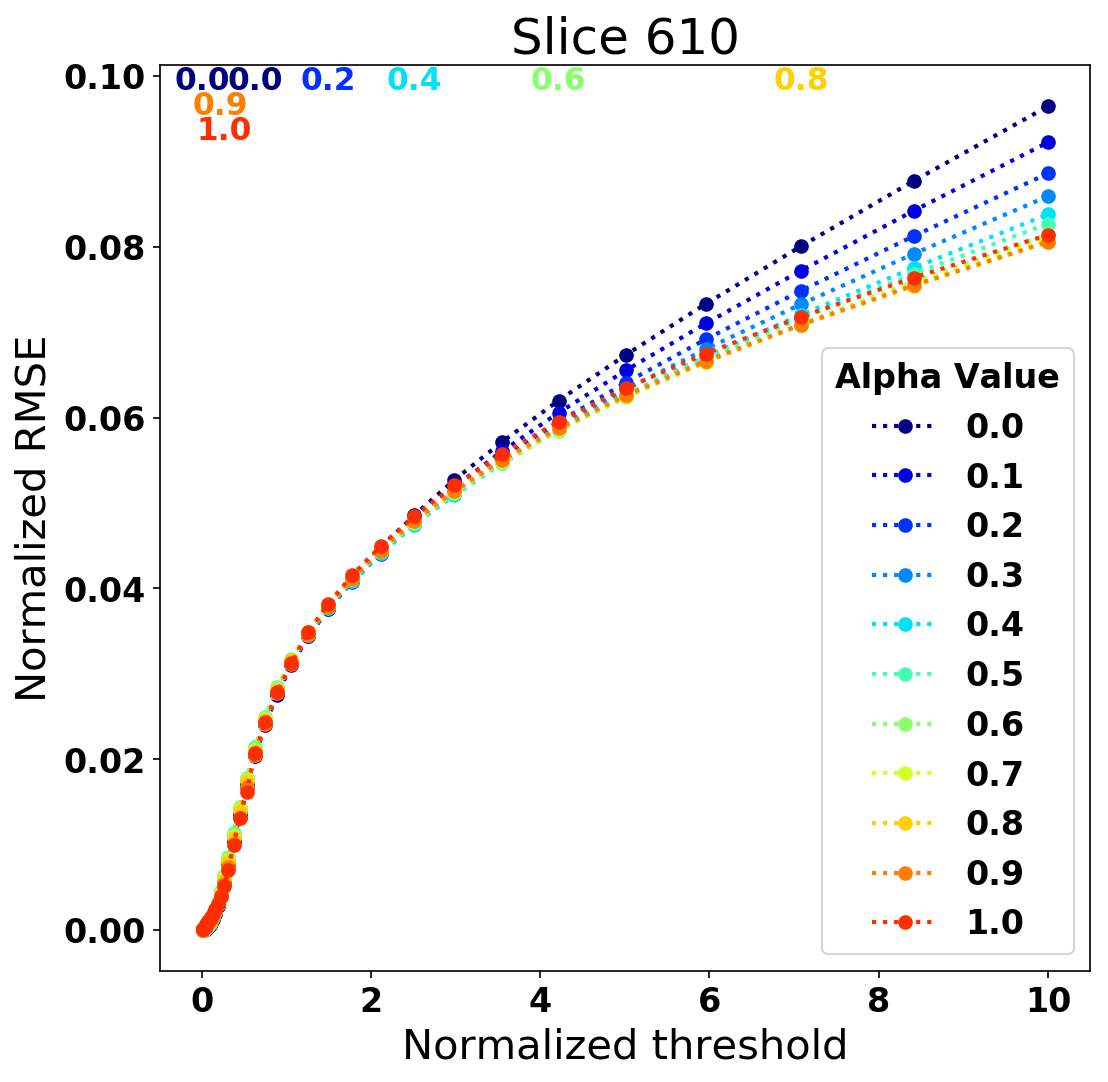}
    \includegraphics[width=120pt]{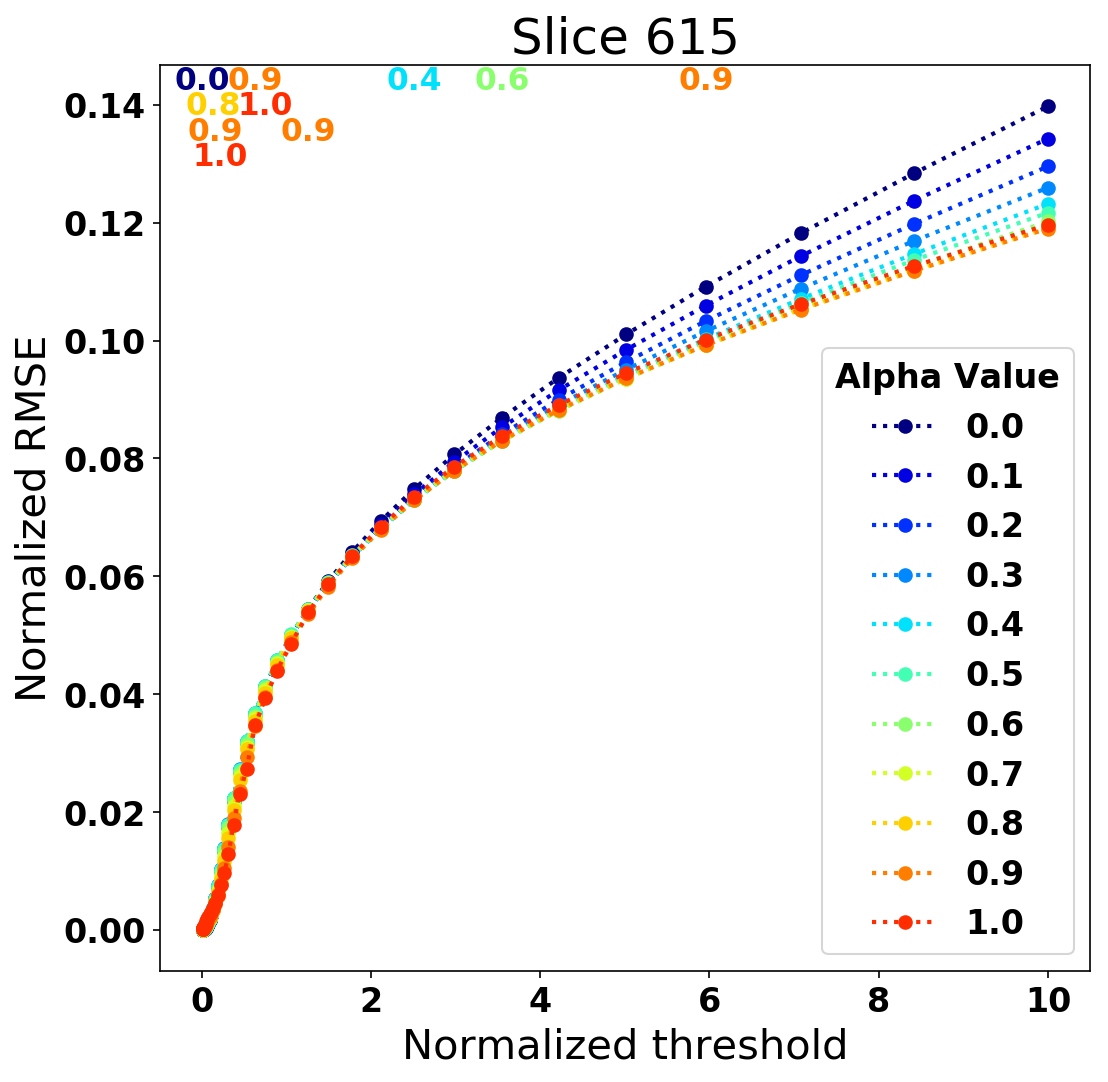}
  \end{center}
  \caption{\label{fig:DMNLA}
           Normalized non-linear approximation curves
           for four different slices of the dark matter distribution.
           For each threshold, the $\alpha$ value corresponding to
           the lowest approximation error is displayed on the top.}
\end{figure}

\begin{figure}[h]
  \begin{center}
    \includegraphics[width=120pt]{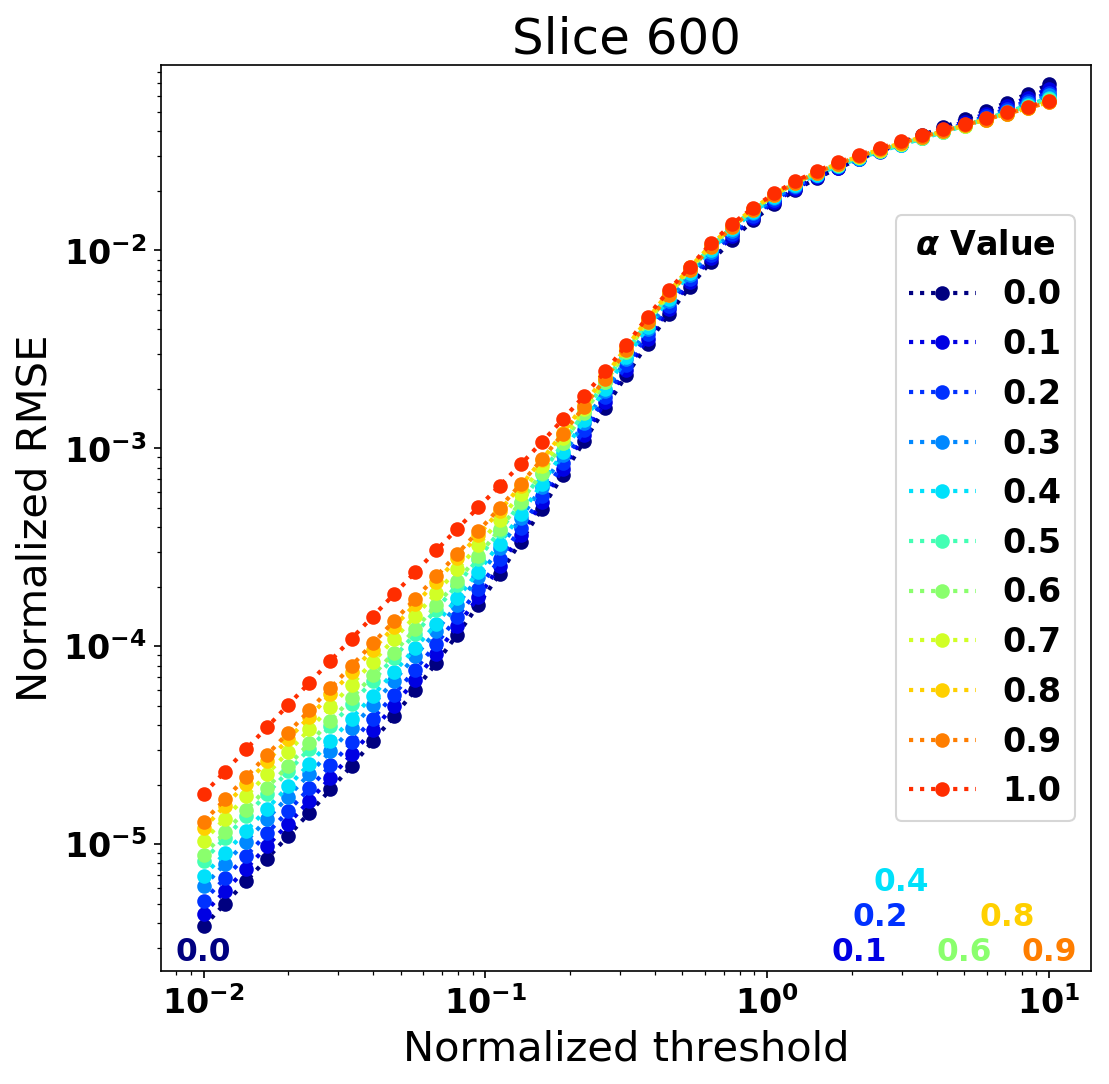}
    \includegraphics[width=120pt]{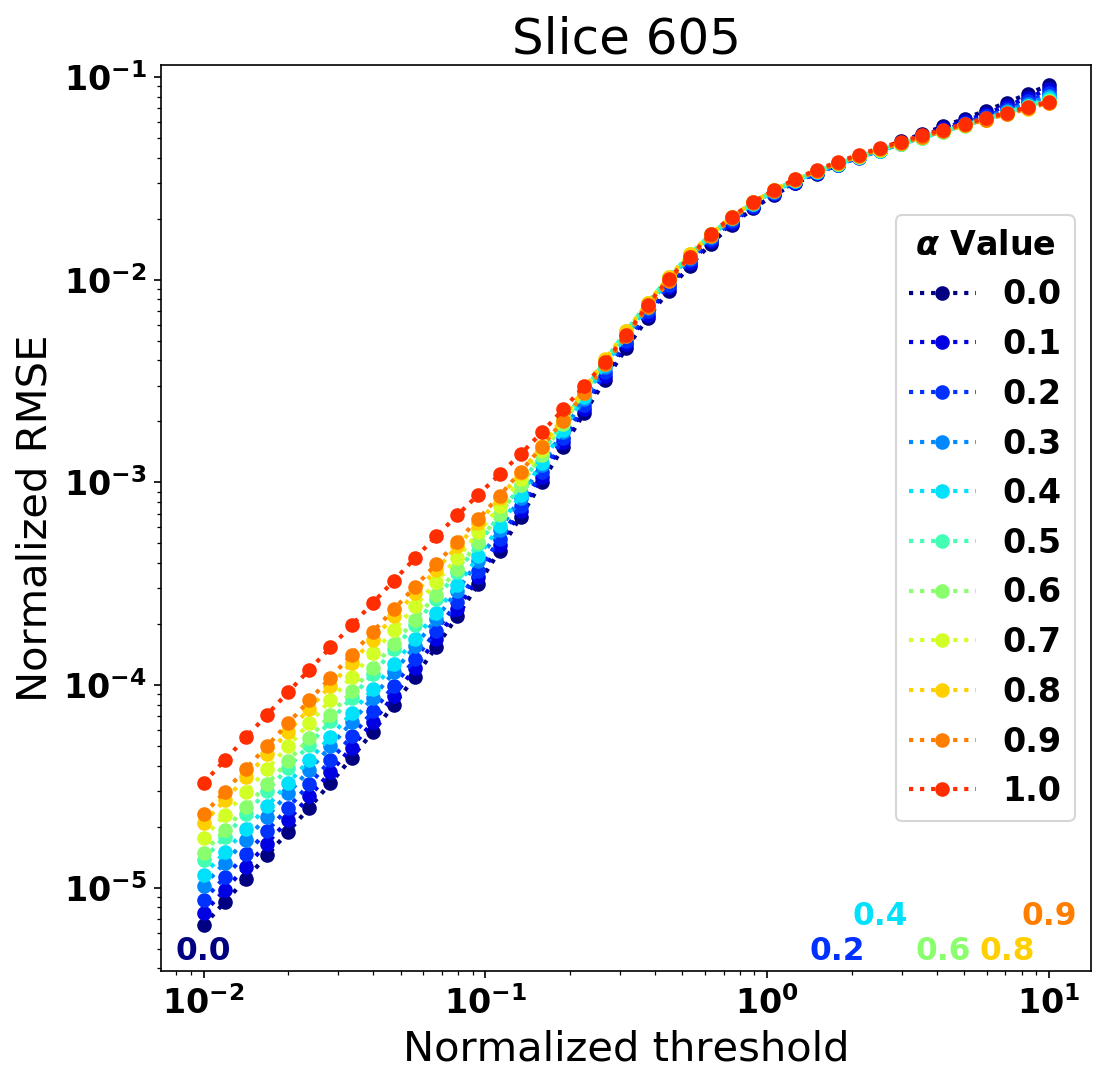}
    \includegraphics[width=120pt]{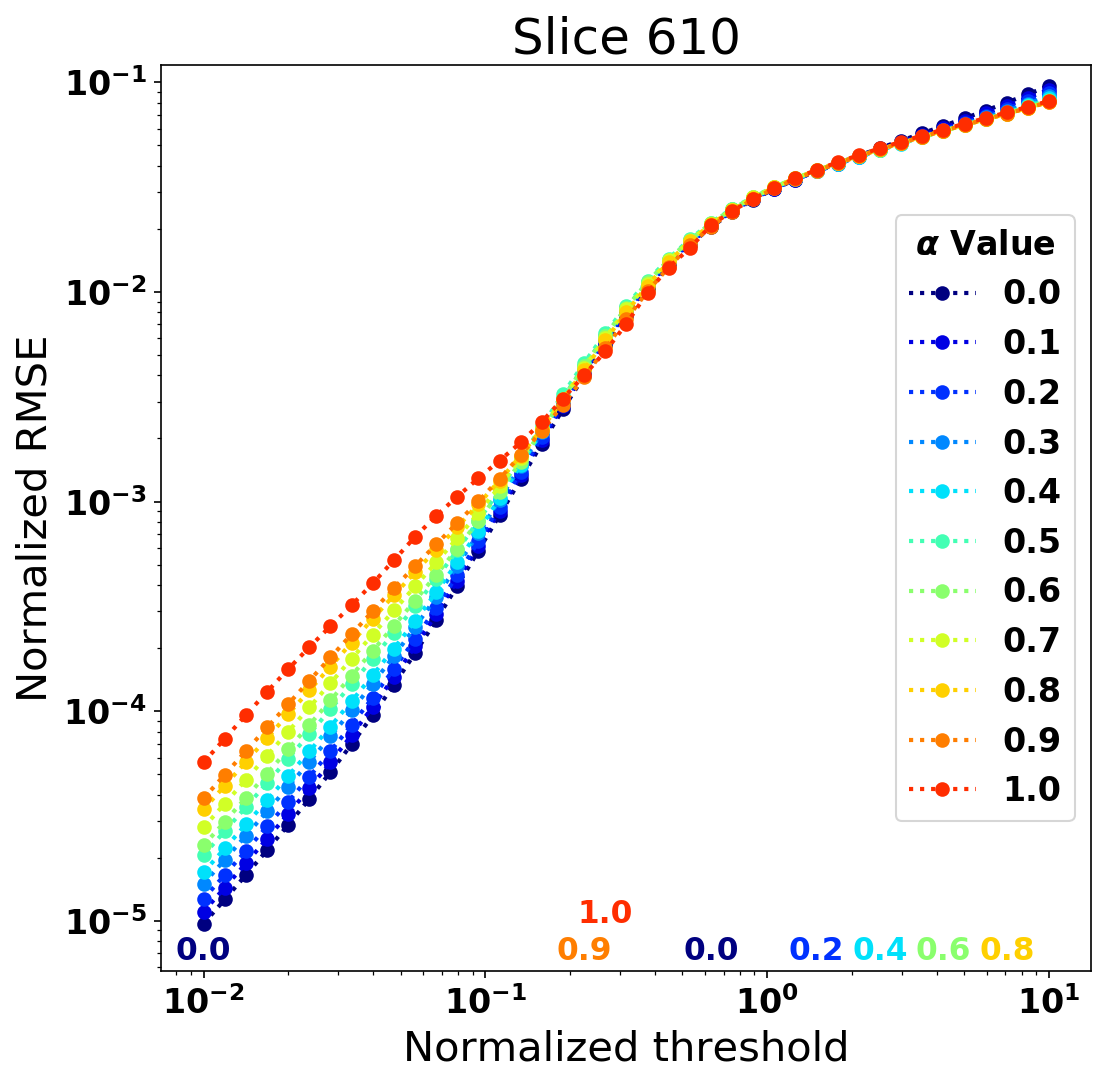}
    \includegraphics[width=120pt]{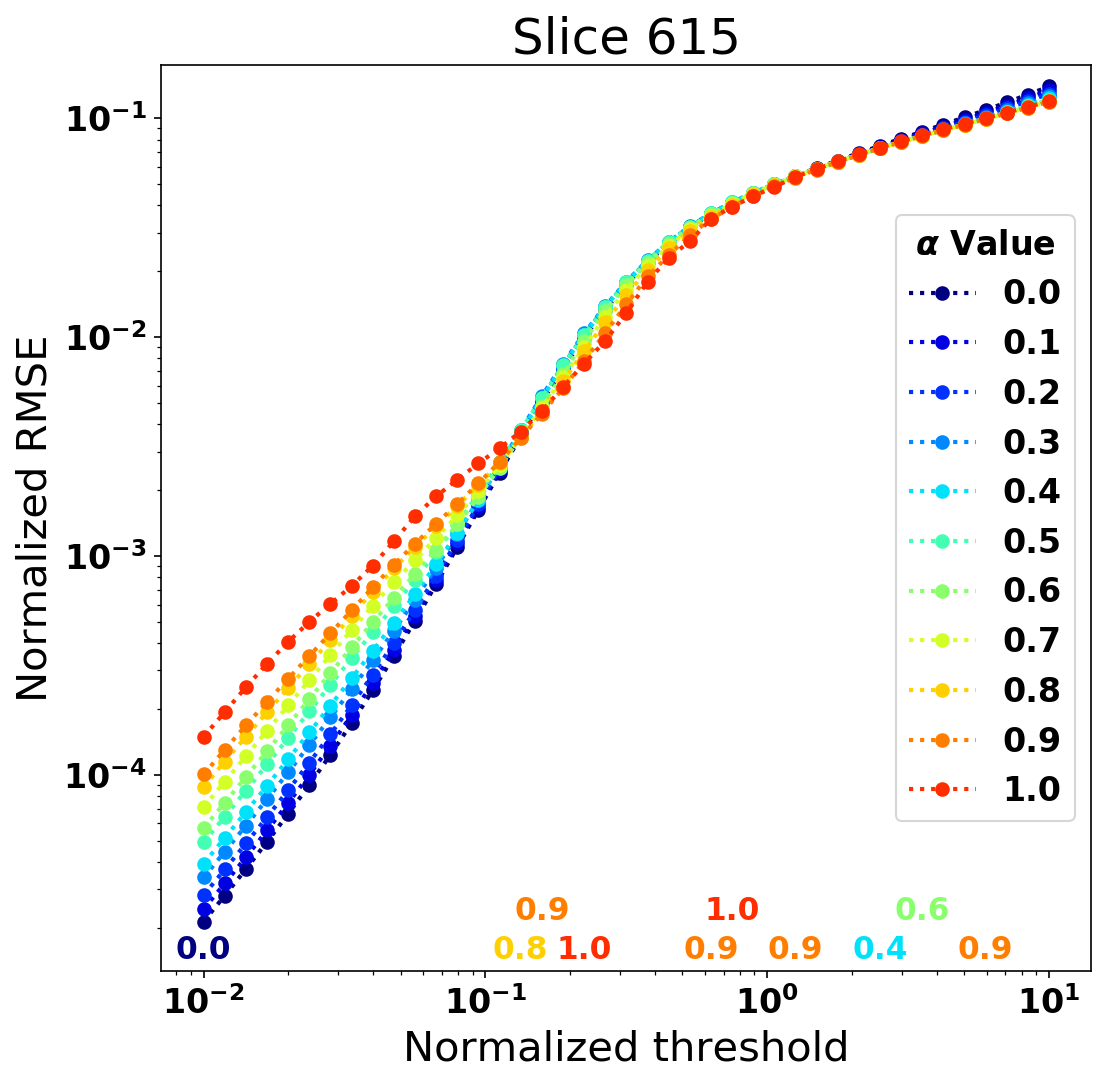}
  \end{center}
  \caption{\label{fig:DMNLA_LOG}
           Normalized non-linear log-approximation curves
           for four different slices of the dark matter distribution.
           For each threshold, the $\alpha$ value corresponding to
           the lowest approximation error is displayed on the bottom.}
\end{figure}

We therefore computed for the patchwork approach
the non-linear approximation curves for the different slices in redshift.
These non-linear approximation curves are
illustrated in linear and log scale in Figs~\ref{fig:DMNLA}
and \ref{fig:DMNLA_LOG}, respectively.
These curves illustrate that for large threshold values,
corresponding to selecting dense core regions,
the $\alpha=0.9$-shearlet system is most suitable. 
For slice $600$ and $605$ (higher redshift), when decreasing the threshold,
there is a transition from $\alpha=0.9$ to $\alpha=0$
(very elongated shearlets) for the best $\alpha$ value.
This can be understood as including more and more filamentary structures
when the threshold decreases.

For lower redshift slices on the other hand,
the best values are obtained more consistently across thresholds for $\alpha=0.9$
or $\alpha=1$ because more core structures and less filaments
are visible in the data.
Overall, this illustrates how adaptive to diverse structures in the data
the $\alpha$-shearlets can be.
Furthermore, it shows that the anisotropy parameter $\alpha$ can be used to
characterize different types of structure present in the data.

\subsection{Computing Requirements}

All codes were run on the same cluster so that we can assess the relative computing time requirements
for the three approaches.

For the rotation-based approach, on the current python implementation using pyFFTW
(\url{https://pypi.org/project/pyFFTW/}) and also based on a
parallelized transform using $6$ cores, denoising a $N_{side} = 2048$ map
using $5$ rotations and $4$ scales of decomposition takes about $35$ minutes
for $\alpha=1$ and $1$ hour for $\alpha=0$ (the most redundant transform). Note that 
time to perform the rotation-based approach scales linearly with the number of rotations. 
In comparison, denoising with the patchwork approach a $N_{side}=2048$ map using $4$ scales of decomposition
(with the same parallelization of the transform as for the rotation-based approach)
takes about $9$ minutes for $\alpha=1$ and $20$ minutes for $\alpha=0$.

For the multiscale dictionary learning algorithm, computing time for the learning phase 
ranged from about $2.5$ hours for scale $0$ to about $3.5$ hours for scale $2$,
when using our C\texttt{++} code with $4$ cores for the sparse coding.
This increase is due to the low value for $\epsilon^{(2)}$ and large value
for the maximal sparsity $K^{(2)}$, even if the training set is smaller than for scale $0$.
Note that learning these dictionaries can be performed in parallel, which was done in practice.
For the dark matter scenario, the learning took about $65$ minutes.

Once the dictionary was learned, sparse coding all patches took typically
from $15$ minutes (scale $2$) to about $22$ minutes (scale $0$) for the thermal
dust map, and $9$ minutes for the dark matter halo distribution, using $24$ cores. 

Overall, the two $\alpha$-shearlet approaches are therefore easier to set up,
with less parameters to optimize that depend directly on the data,
and result in faster denoising than the dictionary learning based approach.

\section{Conclusions}
\label{sect_ccl}

We have proposed two new types of adaptive representations on the sphere:
a patch-based dictionary learning approach and choosing among a parametrized
family of representations, the $\alpha$-shearlets.
To extend these constructs from the Euclidean setting
to data defined on the sphere, we proposed to use overlapping charts
based on the HEALPix framework.
For the dictionary learning technique,
a possible multi-scale extension was presented by learning dictionaries
on each scale after performing a subsampled wavelet decomposition on the sphere.
For the $\alpha$-shearlets, we proposed two approaches to construct the charts:
resampling the sphere according to various rotations
associated with a partition of unity not sensitive to border effects,
or constructing 6 overlapping charts based on composite extended HEALPix faces.

We evaluated all three approaches by conducting denoising experiments on
thermal dust maps, and dark matter maps.

Our main findings are as follows:
\begin{itemize}
  \item[-] thanks to the use of overlapping charts, all of our proposed
           approaches are free of the block artefacts
           that typically appear if one naively uses the disjoint HEALPix
           faces for doing denoising;

  \item[-] in both scenarios investigated,
           the  dictionary learning approach gave the best performance
           by providing atoms adapted to the structure present in the images,
           for a given noise level;

  \item[-] the performance of the dictionary learning approach depends on setting
           several hyper-parameters that depend on the signal observed
           (multiscale or not), and on the training set.
           This approach therefore requires more computing and tuning time
           than the other approaches;

  \item[-] 
           which of the two $\alpha$-shearlet approaches performed better
           depended on the chosen scenario;
           the rotation-based approach involves interpolation
           which is detrimental to capturing signals that vary significantly
           at the scale of just a few pixels,
           but it achieved better results for the thermal dust simulations;

  \item[-] 
           for different values of the anisotropy parameter $\alpha$,
           the $\alpha$-shearlet system is adapted to different structures
           (filaments, dense cores) present in the dark matter halo distribution
           simulation.
\end{itemize}

The respective performance of these approaches depends on the criteria used:
the dictionary learning approach provided the best denoising results
in both scenarios, but has a higher number of parameters to set 
and requires more computing time;
among the $\alpha$-shearlets, the rotation-based approach is best
for smooth signals, but the converse is true for signals
with significant variation at the scale of a few pixels. 
The three proposed approaches can therefore be used
to process data living on the sphere, and choosing the "best" approach 
will depend on the scenario considered as well as the computing resources available.

\section*{\label{sec:repres}Reproducible Research}

In the spirit of reproducible research, we make public our codes on the sphere 
on the common repository \url{github.com/florentsureau/ARES}. The dictionary learning
 and alpha-shearlets codes on the sphere are associated with tutorial jupyter notebooks
 illustrating how to use them for denoising.

\begin{acknowledgements}

  This work is funded by the DEDALE project, contract no. 665044, 
  within the H2020 Framework Program of the European Commission.
  The authors thank the Horizon collaboration
  for making their simulations available.
\end{acknowledgements}

\bibliographystyle{aa} 
\bibliography{refs,references} 

\begin{appendix}

\section{Review of Euclidean $\alpha$-shearlets}
\label{appendix:alpha}

$\alpha$-shearlets are a family or representations
that generalizes wavelets and shearlets.
Like shearlets---originally introduced in
\citet{ShearletsNotConeAdaptedFirstPaper,ConeAdaptedShearletFirstPaper}---they
are a directionally sensitive multiscale system in $\R^2$
improving upon wavelets when it comes to handling data that is
governed by directional features like edges.

\begin{figure}[h]
  \begin{center}
    \includegraphics[width=50pt]{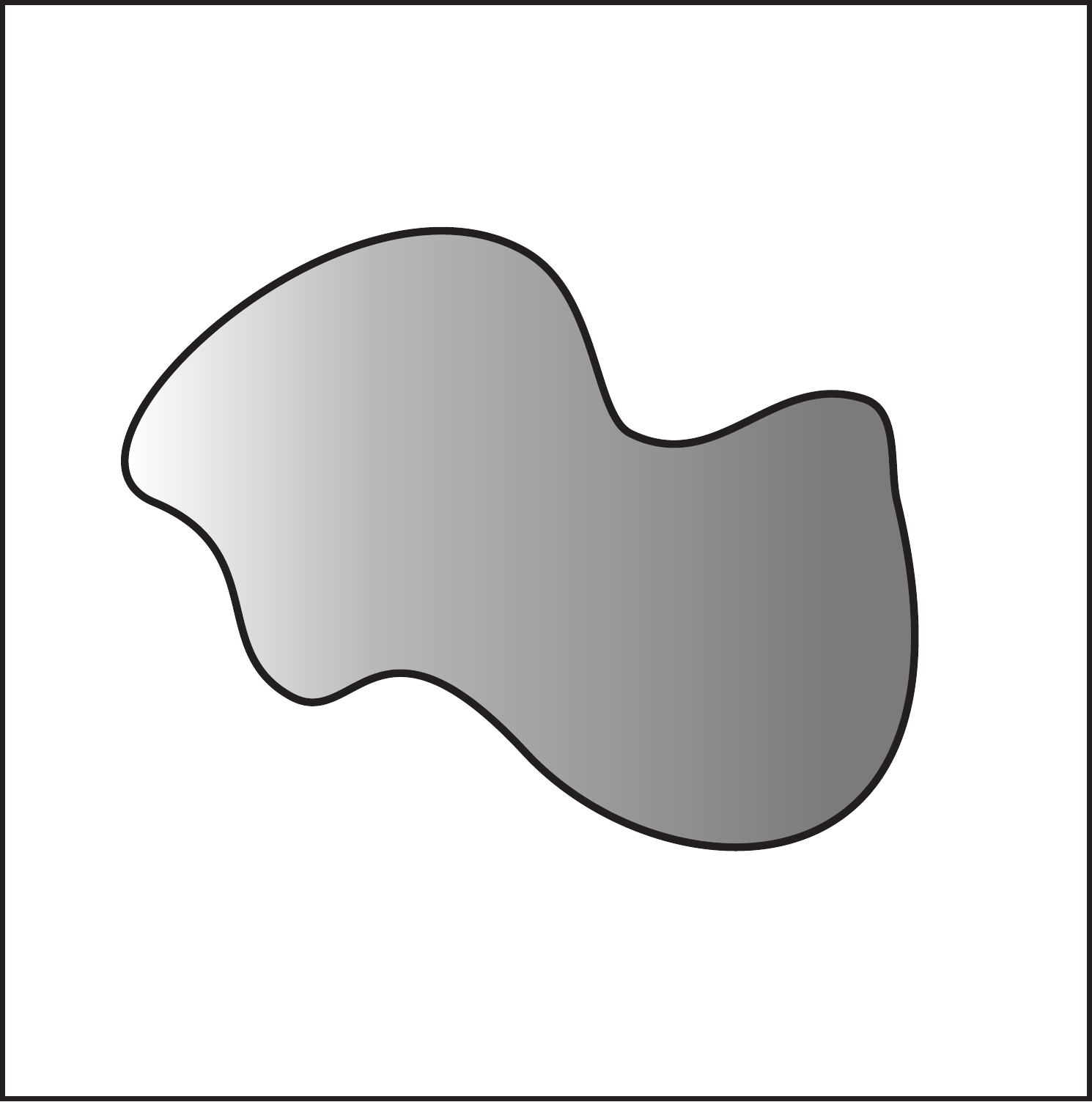}
  \end{center}
  \caption{\label{fig:CartoonLikeFunction}
           An example of a \emph{cartoon-like function}. Such a function $f$
           is smooth, apart from a jump continuity along a curve $\gamma$.
           Even though $f$ might be discontinuous along $\gamma$,
           the boundary curve $\gamma$ itself is required to be smooth.}
\end{figure}

$\alpha$-shearlets are characterized by an \emph{anisotropy parameter}
$\alpha \in [0,1]$, and were designed to yield optimally sparse representations
for the class of \emph{$C^\beta$-cartoon-like functions}
\citep{shearlet_book,
       CompactlySupportedShearletsAreOptimallySparse,
       OptimallySparseMultidimensionalRepresentationUsingShearlets,
       StructuredBanachFrames2},
a model class for natural images \citep{CandesDonohoCurvelets}
as illustrated in Fig.~\ref{fig:CartoonLikeFunction}.

In the remainder of this section, we briefly motivate the choice of
$\alpha$-shearlet systems, discuss the most important
mathematical properties of $\alpha$-shearlet systems,
and then comment on the implementation that we used.

\subsection{Motivation} \label{sub:AlphaShearletMotivation}

Before giving a formal definition of ($\alpha$)-shearlet systems,
it is instructive to roughly compare the operations used for their construction
to the ones used for defining wavelet systems
\citep{DaubechiesTenLecturesOnWavelets}.
Recall (see e.g.~\citet{DaubechiesTenLecturesOnWavelets})
that for a \emph{scaling function} $\phi \in L^2 (\R^d)$ and a
mother wavelet $\psi \in L^2 (\R^d)$,
the associated \emph{(discrete) wavelet system} with
sampling density $\delta > 0$ is given by
\[
  \quad
  \mathcal{W} (\phi, \psi; \delta)
  := \left( \phi (\bullet - \delta k ) \right)_{k \in \Z^d}
     \cup \left(
            2^{d j/2} \cdot \psi(2^j \bullet - \delta k )
          \right)_{j \in \N_0, k \in \Z^d} .
\]
In other words, the wavelet system consists of all translates
of the scaling function $\phi$ along the lattice $\delta \Z^d$,
together with certain translates of the \emph{isotropically dilated}
scaling functions $\psi_j := 2^{dj / 2} \, \psi (2^j \bullet)$.
Here, the wavelet $\psi_j$ on the $j$-th scale is translated along the lattice
$\delta \cdot 2^{-j} \Z^d$, which is adapted to the ``size'' of $\psi_j$.

It is crucial to note that even in dimension $d > 1$,
wavelets use the isotropic dilations $x \mapsto 2^j x$
which treat all directions in the same way.
Therefore, wavelet systems are not optimally suited for
representing functions governed by features with different directions.
Admittedly, instead of using one mother wavelet $\psi$,
it is common to employ wavelet systems that use finitely many mother wavelets
$\psi^{(1)}, \dots, \psi^{(N)}$; usually these are obtained
by choosing each $\psi^{(j)}$ as a certain tensor product of
\emph{one-dimensional} scaling functions and mother wavelets.
But such a modified wavelet system is again only able to distinguish
a fixed number of directions, independent of the scale $j$,
and therefore does not admit a satisfactory directional sensitivity.

\begin{figure}[h]
  \begin{center}
    \includegraphics[width=210pt]{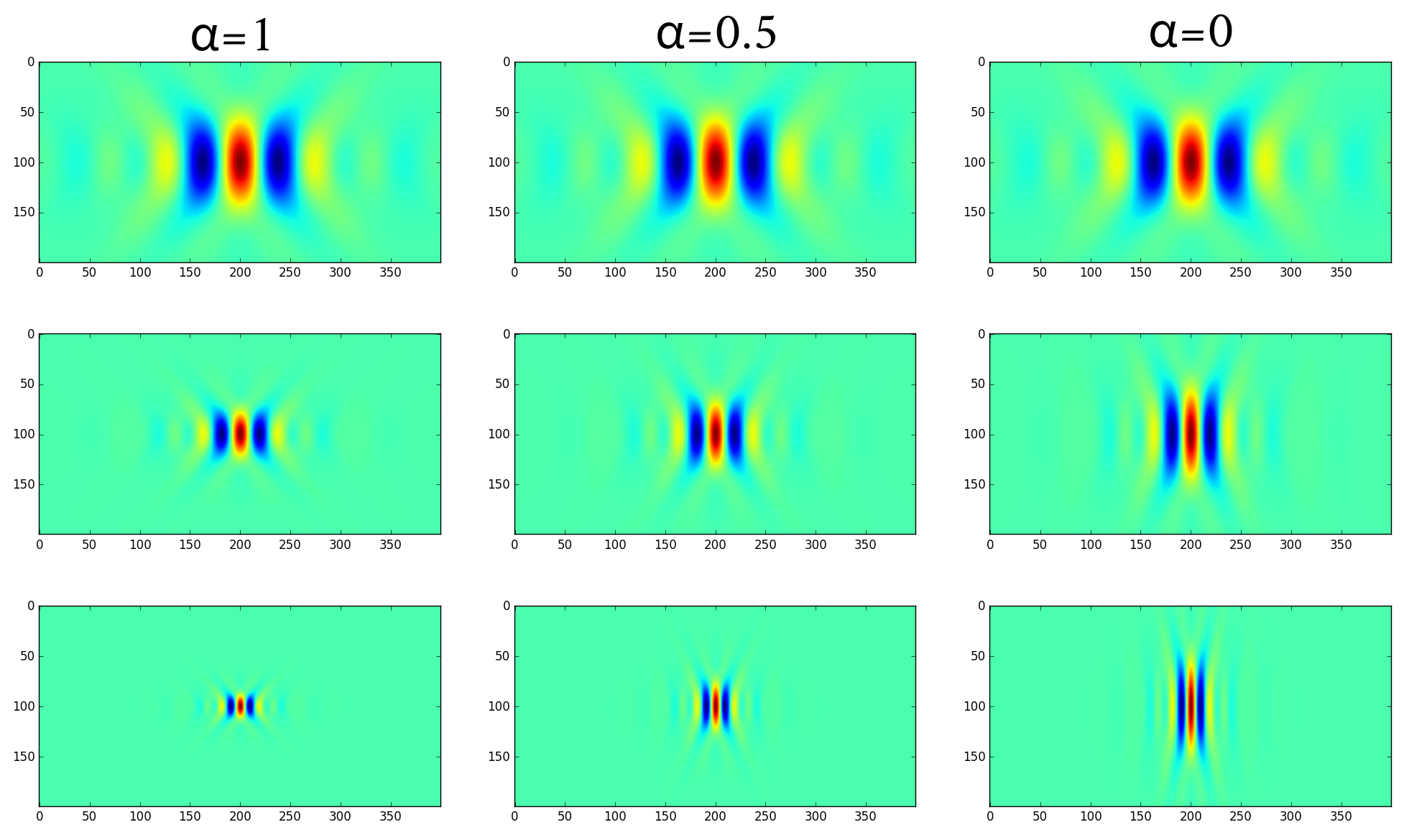}
  \end{center}
  \caption{\label{fig:AlphaEffect}
           The effect of dilating a ``prototype function'' $\psi$
           (shown at the top of each row) with the matrices $D_j^{(\alpha)}$
           to obtain $\psi(D_j^{(\alpha)} \bullet)$,
           for different values of the scale $j$
           (going from $j=0$ (top) to $j=2$ (bottom))
           and of the ``anisotropy parameter'' $\alpha \in [0,1]$.}
\end{figure}

To overcome this problem, shearlets (like curvelets)
use the \emph{parabolic dilation matrices}
$D_j^{(1/2)} := \left(
                  \begin{smallmatrix}
                    2^j & 0       \\
                    0   & 2^{j/2}
                  \end{smallmatrix}
                \right)$.
More generally, $\alpha$-shearlets employ the $\alpha$-parabolic dilation matrices
\[
  \qquad \qquad
  \qquad \qquad
  D_j^{(\alpha)}
  := \left(
       \begin{matrix}
         2^j & 0            \\
         0   & 2^{\alpha j}
       \end{matrix}
     \right)
  \quad \text{for} \quad j \in \N_0 \, .
\]
As shown in Fig.~\ref{fig:AlphaEffect}, dilating a function $\psi$ with these
matrices $D_j^{(\alpha)}$ produces functions
$\psi_j^{(\alpha)} = \psi(D_j^{(\alpha)} \bullet)$
which are more elongated along the $x_2$-axis than along the $x_1$-axis,
where the anisotropy is more pronounced for larger values of $\alpha$ or $j$.
The support of the dilated function
satisfies $2^{-j\alpha} \approx \mathrm{height} \approx \mathrm{width}^\alpha$.

It is apparent from Fig.~\ref{fig:AlphaEffect} that for $\alpha < 1$
and large $j \in \N_0$, the functions $\psi_j^{(\alpha)}$ have a
distinguished \emph{direction}.
More precisely, if (as in the figure) $\psi$ oscillates along the $x_1$-axis,
then $\psi_j^{(\alpha)}$ is similar to a \emph{sharp jump along the $x_2$-axis}.
Since we want our dictionary to be able to represent jumps along
\emph{arbitrary} directions, we have to allow some way of changing
the direction of the elements $\psi_j^{(\alpha)}$.
The most intuitive way for achieving this is to use \emph{rotations},
as was done in the construction of
(second generation) curvelets \citep{CandesDonohoCurvelets}.
But later on, it was noted in
\citet{ShearletsNotConeAdaptedFirstPaper,ConeAdaptedShearletFirstPaper}
that from an implementation point of view, rotations have the disadvantage
that they do not leave the digital grid $\Z^2$ invariant.
Therefore, instead of rotations, ($\alpha$)-shearlets use
the \emph{shearing matrices}
\[
  \qquad \qquad
  \qquad \qquad \quad
  S_x := \left( \begin{matrix} 1 & 0 \\ x & 1 \end{matrix} \right)
\]
to adjust the direction of the functions $\psi_j^{(\alpha)}$.
Note though that the shearing matrices $S_x$, $x \in (-\infty, \infty)$
can never cause an effect similar to a rotation with angle
$\theta$ for $|\theta| > 90\degree$.
Therefore, for the definition of a \emph{cone-adapted shearlet system}, one
only uses shearings corresponding to rotations with angle
$|\theta| \leq 45\degree$, and then uses a modified mother shearlet
$\psi^\natural$ to cover the remaining directions.

Collecting all previously described constructs,
the cone-adapted $\alpha$-shearlet system
with sampling density $\delta > 0$,
associated to a low-pass filter $\varphi \in L^2 (\R^2)$,
and mother shearlet $\psi \in L^2 (\R^2)$ is defined as:
\begin{equation}
  \label{defn:AlphaShearlets}
  \begin{split}
    \mathrm{SH}_\alpha (\varphi, \psi ; \delta)
    & := \left( \varphi (\bullet - \delta k) \right)_{k \in \Z^2} \\
    & \quad \cup \left(
                   2^{(1 + \alpha) j / 2} \,
                   \psi (R^\iota D_j^{(\alpha)} S_\ell \bullet - \delta k)
                 \right)_{(j,\ell, \iota) \in I, k \in \Z^2} \, ,
  \end{split}
\end{equation}
with $R := \left( \begin{smallmatrix} 0 & 1 \\ 1 & 0 \end{smallmatrix} \right)$,
and
\[
  \qquad
  I := I^{(\alpha)}
    := \left\{
         (j,\ell,\iota) \in \N_0 \times \Z \times \{0,1\}
         \, : \,
         |\ell| \leq \lceil 2^{j(1-\alpha)} \rceil
       \right\} \, .
\]

For brevity, let us set
$\psi_{j,\ell,\iota}^{(\alpha)}
 := 2^{(1+\alpha) j / 2} \, \psi \left( R^{\iota} \, D_{j}^{(\alpha)} \, S_\ell \, \bullet \right)$,
and observe with this notation that
\begin{equation}
  \qquad
  2^{(1 + \alpha) j / 2} \,
  \psi
  \left(
    R^{\iota} \, D_{j}^{(\alpha)} \, S_\ell \,  \bullet - \delta k
  \right)
  = \psi_{j,\ell,\iota}^{(\alpha)}
    \left( \bullet - \delta A_{j,\ell,\iota}^{-1} k \right),
  \label{eq:AlphaShearletTranslationClarification}
\end{equation}
with $A_{j,\ell,\iota} := R^{\iota} \, D_{j}^{(\alpha)} \, S_\ell$.

\subsection{Mathematical properties} \label{sub:AlphaShearletsMathematicalProperties}

The most basic property of $\alpha$-shearlets that we will be interested in
is that they indeed form a \emph{(redundant) representation system}
for $L^2 (\R^2)$.
In mathematical terms, this means that the $\alpha$-shearlet system forms a
\emph{frame} \citep{ChristensenFramesAndRieszBases},
for a suitable choice of the generators $\varphi,\psi$.
In particular, if  $\varphi, \psi \in L^2(\R^2)$ have compact support and
satisfy certain decay and smoothness conditions
(see \citet[Theorem 5.10]{StructuredBanachFrames2} for details),
then there is a ``minimal sampling density'' $\delta_0 > 0$,
such that the $\alpha$-shearlet system is indeed a frame for $L^2 (\R^2)$,
for all $0 < \delta \leq \delta_0$.

The main motivation for introducing ($\alpha$)-shearlets
was the wish for a representation system which is better
adapted to data governed by directional features,
which are often present in natural images, and also in astronomical images.
One key result relates ($\alpha$)-shearlets to $C^{1/\alpha}$-cartoon-like functions.

Roughly speaking, a function $f \in L^2 (\R^2)$ is called a
\emph{$C^\beta$-cartoon-like function}, written $f \in \mathcal{E}^\beta (\R^2)$
(with $\beta \in (1,2]$), if $f = f_1 + f_2 \cdot \Indicator_B$ for certain
$f_1, f_2 \in C_c^\beta ([0,1]^2)$ and such that the set $B \subset [0,1]^2$
has a boundary curve of regularity $C^\beta$.
For a more formal definition, we refer to
\citet[Definition 6.1]{StructuredBanachFrames2}.

Using this notion, we have the result that the best $N$-term approximation
error with such a frame of $\alpha$-shearlets
(that is, the smallest approximation error obtained by a linear combination
of $N$ $\alpha$-shearlets)
is decaying at (almost) the best rate that any dictionary $\Psi$
can reach for $C^{\beta}$-cartoon-like functions;
see \citet[Theorem 6.3]{StructuredBanachFrames2}
for a more precise formulation of this result.
To obtain this optimal approximation rate, the anisotropy parameter $\alpha$
needs to be adapted to the regularity $\beta$
of the $C^\beta$-cartoon-like functions, that is, $\alpha = 1/\beta$.
In general, given a certain data set, or a certain data model,
different types of $\alpha$-shearlet systems will be better adapted
to the given data than other $\alpha'$-shearlet systems.
In Section \ref{sect_exp}, we will verify this for specific sets
of data living on the sphere.

\begin{figure}[h]
  \begin{center}
    \includegraphics[width=110pt]{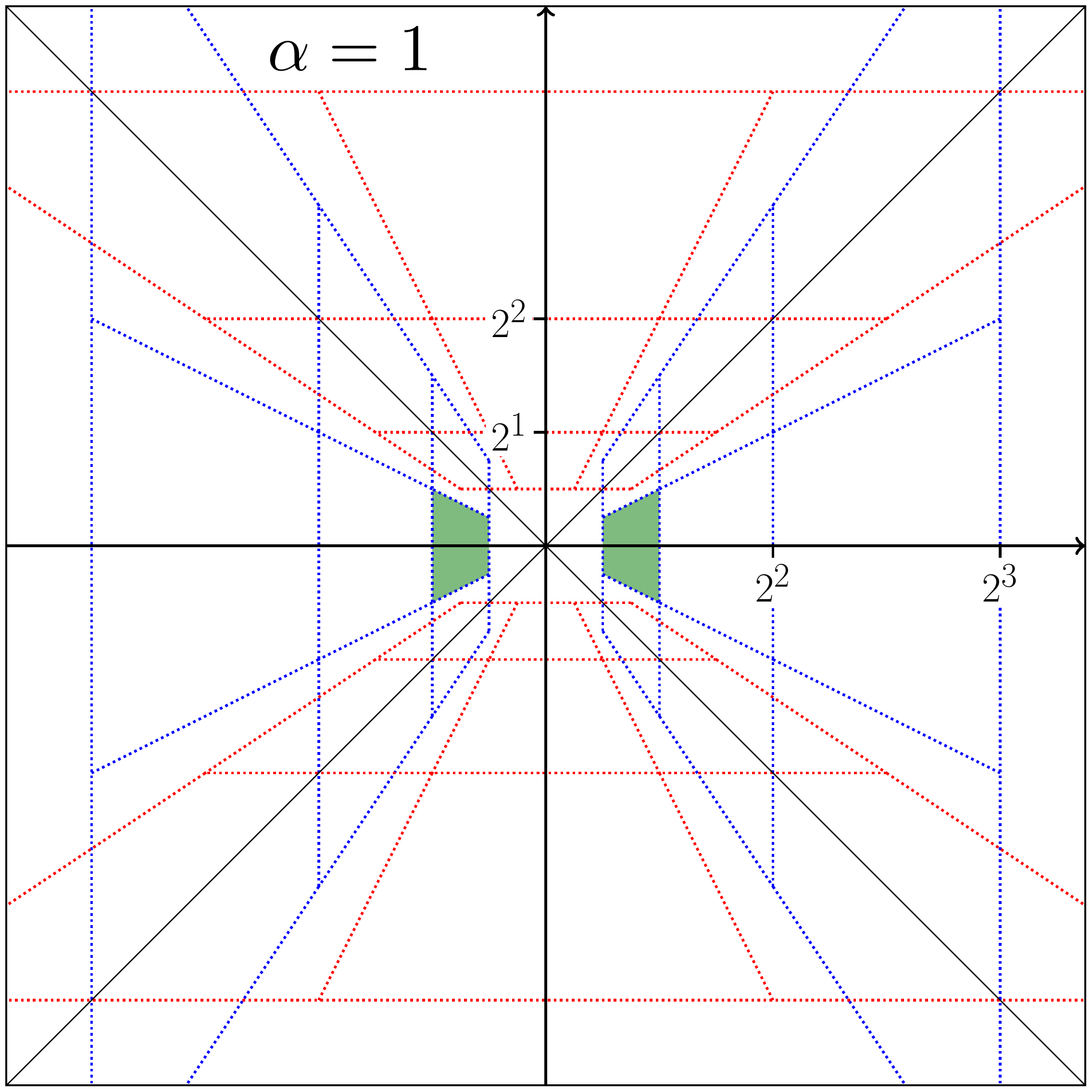}
    \includegraphics[width=110pt]{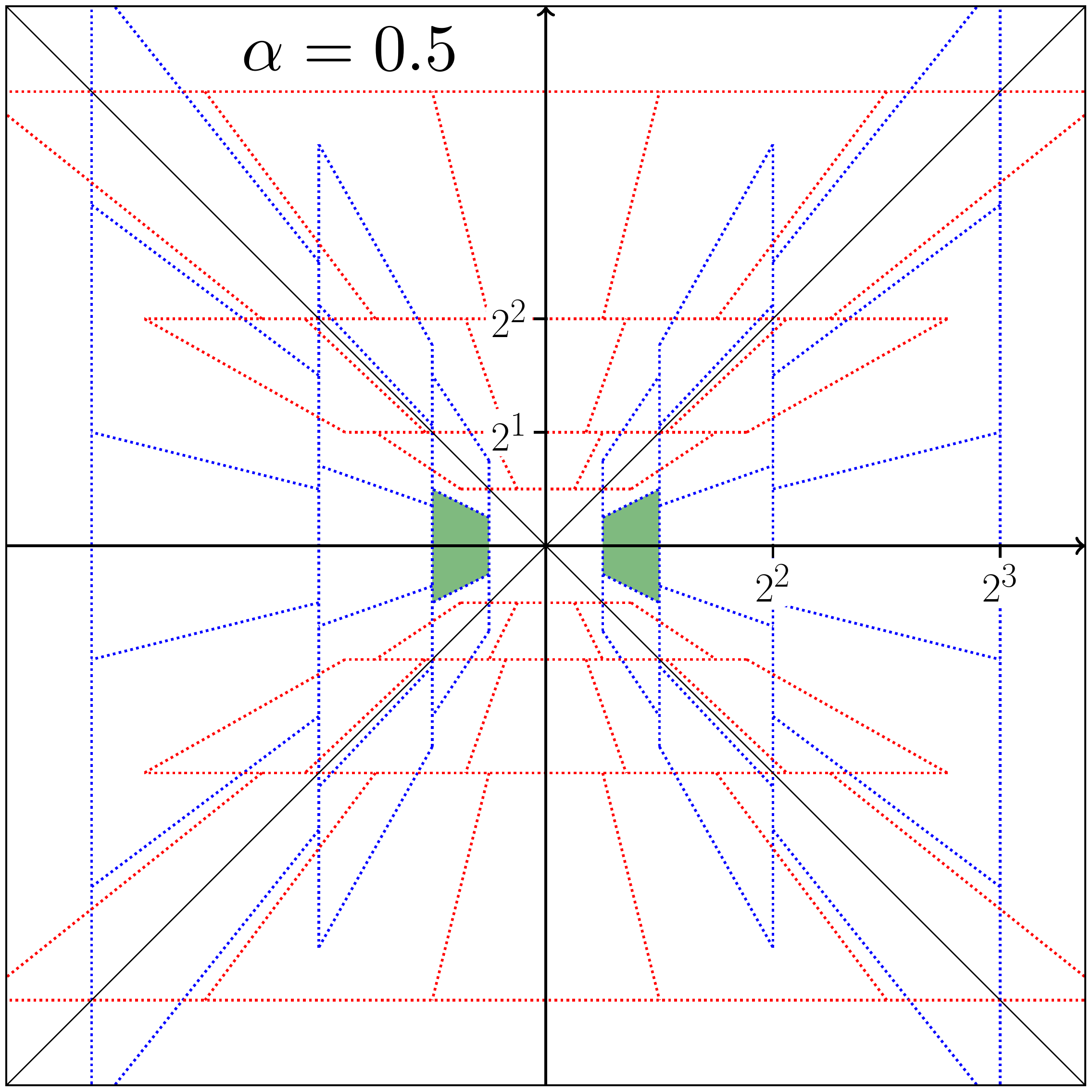}
    \includegraphics[width=110pt]{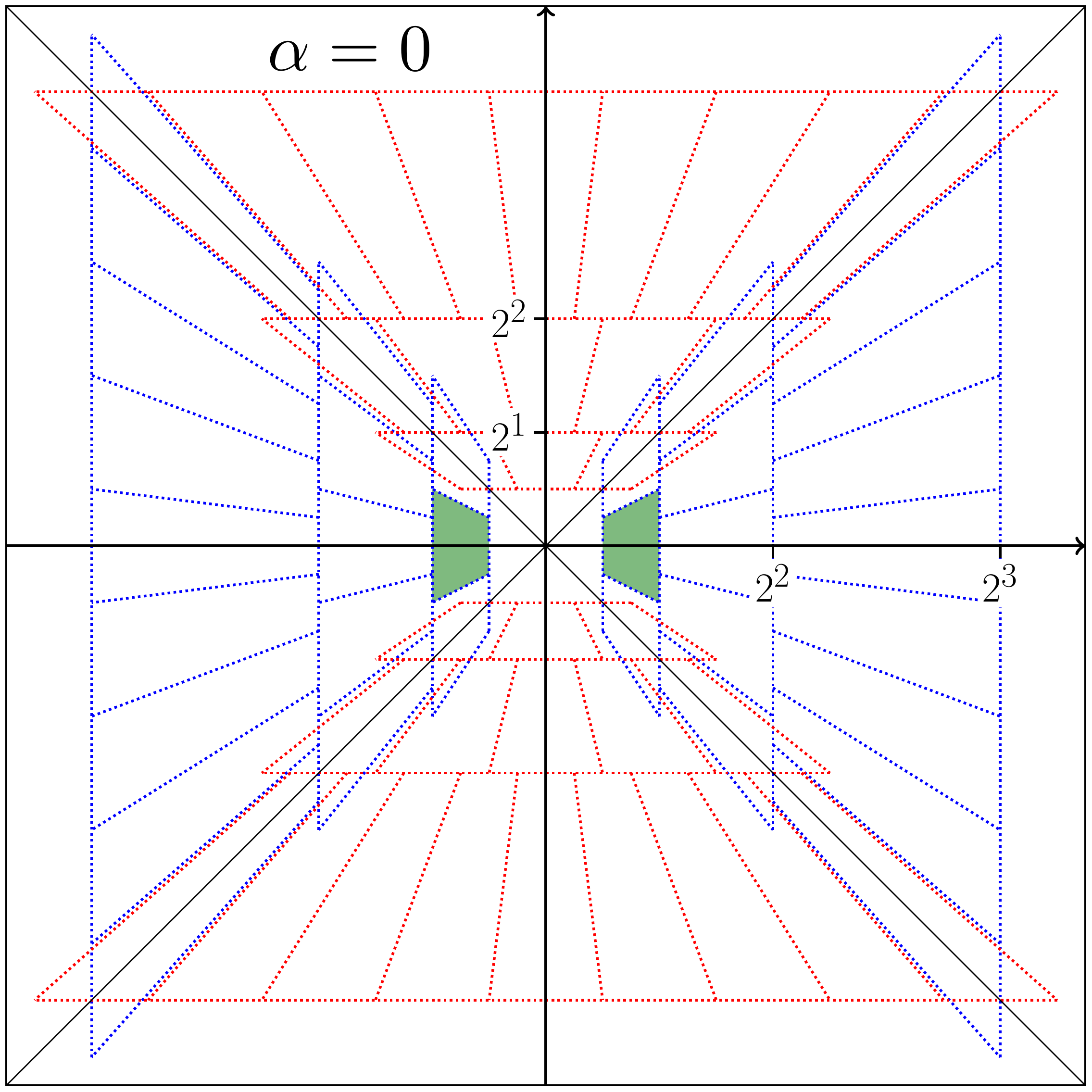}
  \end{center}
  \caption{\label{fig:AlphaShearletFrequencyConcentration}
           The frequency concentration of $\alpha$-shearlets for
           different values of $\alpha$.
           One sees that each ``dyadic annulus'' $\{\xi \,:\, |\xi| \asymp 2^{j}\}$
           is split into a number $N_j^{(\alpha)}$ of ``wedges''
           representing the different directions.
           In fact, $N_j^{(\alpha)} \asymp 2^{(1-\alpha) j}$.}
\end{figure}

We close our discussion of the mathematical properties of
$\alpha$-shearlet systems with a brief discussion of the
\emph{frequency concentration} of such systems.
To this end, assume for the moment that the ``mother shearlet'' $\psi$
is concentrated in frequency to the set
\[
  Q
  := \{
       \xi \in \R^2
       \, : \,
       3^{-1} \leq |\xi_1| \leq 3 \text{ and } |\xi_2| \leq |\xi_1|
     \} \, ,
\]
which is a union of two opposing ``wedges''
(highlighted in green in Fig.~\ref{fig:AlphaShearletFrequencyConcentration}).
From elementary properties of the Fourier transform,
one then sees that each $\alpha$-shearlet $\psi_{j,\ell,\iota}^{(\alpha)}$
has frequency support in $S_\ell^T D_j^{(\alpha)} R^{\iota} Q$,
where we denote by $A^T$ the transpose of a matrix $A$.
The resulting coverings of the frequency plane for different values
of the anisotropy parameter $\alpha$ are shown in
Fig.~\ref{fig:AlphaShearletFrequencyConcentration}.

Together, Figs~\ref{fig:AlphaEffect} and
\ref{fig:AlphaShearletFrequencyConcentration} show that the
parameter $\alpha$ has three different, but related effects:
\begin{itemize}
  \item It affects the ``shape'' of the elements of the $\alpha$-shearlet system.
        Indeed, Fig.~\ref{fig:AlphaEffect} shows that
        $\mathrm{height} \approx \mathrm{width}^\alpha$.

  \item It affects the directional selectivity:
        As seen in Fig.~\ref{fig:AlphaShearletFrequencyConcentration},
        on scale $j$, an $\alpha$-shearlet system can distinguish about
        $2^{(1-\alpha)j}$ different directions.

  \item It affects the frequency support of the elements of the
        $\alpha$-shearlet system, see
        Fig.~\ref{fig:AlphaShearletFrequencyConcentration}.
\end{itemize}

\subsection{Implementation} \label{sub:AlphaShearletsImplementation}
The git repository of our implementation of the Euclidean $\alpha$-shearlet
transform can be found at \url{github.com/dedale-fet/alpha-transform},
with extensive documentation available at
\url{rawgit.com/dedale-fet/alpha-transform/master/build/html/AlphaTransform.html}.
Our software package is implemented in Python3 \citep{PythonTutorial},
using NumPy \citep{NumPyArray}.

In this section, we give a rough overview over what the transform computes,
and how it can be used.
Our software package implements two different versions of the $\alpha$-shearlet
transform: A \emph{fully-sampled (non-decimated)} version,
and a \emph{subsampled (decimated)} version.
For the fully-sampled version, the computed coefficients are the (discrete)
convolutions $\varphi \ast f$ and $\psi_{j,\ell,\iota}^{(\alpha)} \ast f$
(for a certain range of scales $j=0,\dots,j_{\max}$),
where the filters $\varphi$ and $\psi_{j,\ell,\iota}^{(\alpha)}$
are chosen as in Equations~\eqref{defn:AlphaShearlets}
and \eqref{eq:AlphaShearletTranslationClarification}.
Thus, for a given input image $f \in \CC^{N \times N}$,
the resulting coefficients form a three-dimensional tensor
of dimension $N_{\alpha, j_{\max}} \times N \times N$,
where
the integer $N_{\alpha, j_{\max}}$ is the total number
of $\alpha$-shearlet filters that is used,
and where each $N \times N$ component of the tensor
is the discrete convolution of $f$ with one of the $\alpha$-shearlet filters.
When considering $j_{\max}$ many scales (i.e., $j = 0,\dots, j_{\max} - 1$)
and if $\alpha < 1$, then
\begin{equation}
  \,\,
  N_{\alpha, j_{\max}}
  = 1 + 2 \cdot \!\! \sum_{j = 0}^{j_{\max} - 1} \!\!
                                \# \{
                                     -\lceil 2^{(1-\alpha)j} \rceil,
                                     \dots,
                                     \lceil 2^{(1-\alpha)j} \rceil
                                   \}
  \asymp 2^{(1-\alpha) j_{\max}} \, .
  \label{eq:FullySampledRedundancy}
\end{equation}
In particular, for $\alpha = 0$, note $N_{0, j_{\max}} \asymp 2^{j_{\max}}$,
so that the redundancy of the fully sampled $\alpha$-shearlet frame
grows very quickly when increasing the number of scales.

To motivate the subsampled transform, note that according to
Eq.~\eqref{defn:AlphaShearlets},
the $\alpha$-shearlet system does \emph{not} contain all translations
of the functions $\varphi$ and $\psi_{j,\ell,\iota}^{(\alpha)}$.
Rather, $\varphi$ is shifted along the lattice $\delta \Z^2$, and---as seen in
Eq.~\eqref{eq:AlphaShearletTranslationClarification}---$\psi_{j,\ell,\iota}^{(\alpha)}$
is shifted along the lattice $\delta A_{j,\ell,\iota}^{-1} \Z^2$,
with $A_{j,\ell,\iota} = R^{\iota} \, D_{j}^{(\alpha)} \, S_\ell$.
Effectively, this means that the full convolution
$f \ast \psi_{j,\ell,\iota}^{(\alpha)}$ is only sampled at certain points,
where the sampling density gets more dense as the scale $j$ increases.
The subsampled version of the $\alpha$-shearlet transform
computes these coefficients.
Internally, this is achieved by using the ``frequency wrapping'' approach
outlined in \citet[Sections 3.3 and 6]{FastDiscreteCurvelet},
\citet[Chapter 4]{ArnaudPhD}, and \citet{LowRedundancyCurveletTransform}
for the case of the curvelet transform.
Since each convolution is sampled along a different lattice,
the subsampled transform of a given image
$f$ is \emph{a list of rectangular matrices of varying dimension}.
This will become more clear in the example below.
One can show for the subsampled transform that the total number
$M = M(\alpha, j_{\max}, N)$ of $\alpha$-shearlet
coefficients for an $N \times N$ image is bounded,
i.e., $M(\alpha, j_{\max},N) \leq M_0 \cdot N^2$, with $M_0$ independent
of $\alpha, j_{\max}, N$.
This is in stark contrast to the fully sampled transform
(at least for $\alpha < 1$), where the total number of coefficients is
$\approx 2^{(1-\alpha) j_{\max}} \cdot N^2$,
see Equation~\eqref{eq:FullySampledRedundancy}.

The main effect of choosing the fully sampled transform is that one gets a
\emph{translation-invariant transform} (i.e., taking the transform of a shifted
image is the same as shifting each component of the coefficient tensor),
and the increased redundancy.
This increased redundancy can actually be beneficial
for certain tasks like denoising, but it can greatly impact
the memory footprint and the runtime:
Computations using the subsampled transform are usually
much faster and require much less memory, but yield slightly worse results.

We close this section with a short IPython session
showing how our implementation of the $\alpha$-shearlet transform
can be used.

\begin{lstlisting}
>>> # Importing necessary packages
>>> from AlphaTransform import AlphaShearletTransform as AST
>>> import numpy as np; from scipy import misc

>>> im = misc.face(gray=True);  im.shape
(768, 1024)

>>> # Setting up the transform.
>>> trafo = AST(im.shape[1], im.shape[0], [0.5]*3, subsampled=False, verbose=False, real=True) # 1

>>> # Computing the alpha-shearlet coefficients
>>> coeff = trafo.transform(im); print(type(coeff)); print(coeff.shape) # 2
<class 'numpy.ndarray'>
(27, 768, 1024)
>>> trafo.indices # 3
[-1,
 (0, -1, 'h'), (0, 0, 'h'), (0, 1, 'h'),
 (0, 1, 'v'), (0, 0, 'v'), (0, -1, 'v'),
 (0, -1, 'l'), (0, 0, 'l'), (0, 1, 'l'),
 (1, -2, 'h'), (1, -1, 'h'), (1, 0, 'h'), ... ]

>>> recon = trafo.inverse_transform(coeff) # 4
>>> np.allclose(recon, im)
True

>>> # Setting up the subsampled transform.
>>> trafo2 = AST(im.shape[1], im.shape[0], [0.5]*3, subsampled=True, verbose=False, real=False) # 5

>>> # Computing the subsampled alpha-shearlet coefficients
>>> coeff2 = trafo2.transform(im); print(type(coeff2)); print(type(coeff2[0])); print(coeff2[0].shape); print(coeff2[1].shape) # 6
<class 'list'>
<class 'numpy.ndarray'>
(129, 129)
(364, 161)
>>> trafo2.indices # 7
[-1,
 (0, -1, 'r'), (0, 0, 'r'), (0, 1, 'r'),
 (0, 1, 't'), (0, 0, 't'), (0, -1, 't'),
 (0, -1, 'l'), (0, 0, 'l'), (0, 1, 'l'),
 (0, 1, 'b'), (0, 0, 'b'), (0, -1, 'b'),
 (1, -2, 'r'), (1, -1, 'r'), (1, 0, 'r'), ... ]
>>> recon2 = trafo2.inverse_transform(coeff2); np.allclose(recon2, im)
True
>>> print(trafo.redundancy); print(trafo2.redundancy) # 8
27
12.08676528930664
\end{lstlisting}

In the line marked with $\# 1$, we set up the $\alpha$-shearlet transform
object \lstinline{trafo}.
Roughly speaking, this will precompute all necessary $\alpha$-shearlet filters,
which are stored in the \lstinline{trafo} object.
The first two parameters of the constructor simply determine the shape
of the images for which the \lstinline{trafo} object can be used,
while the third parameter determines the number of scales $j_{\max}$ to be used,
as well as the value of the anisotropy parameter $\alpha$.
Passing \lstinline{[alpha_0] * N} will construct an
$\alpha$-shearlet transform with $N$ scales (plus the low-pass)
and with $\alpha$ given by \lstinline{alpha_0}.
The \lstinline{verbose} parameter simply determines how much additional output
(like a progress bar) is displayed.
The \lstinline{subsampled} parameter determines whether the non-decimated,
or the decimated transform is used.
Finally, the \lstinline{real} parameter determines whether real-valued or
complex-valued $\alpha$-shearlet filters are used.
Essentially, real-valued filters have frequency support
in the union of two opposing wedges
(as shown in Fig.~\ref{fig:AlphaShearletFrequencyConcentration}),
while for complex-valued filters, one gets two filters for each real-valued one:
one complex-valued filter has frequency support in the ``left'' wedge,
while the other one is supported in the ``right'' wedge.

In line $\# 2$, we use the \lstinline{transform()} method of the constructed
\lstinline{trafo} object to compute the $\alpha$-shearlet transform of
\lstinline{im}.
As seen, the result is an ordinary NumPy array of dimension
${N_{\alpha, j_{\max}} \times N_1 \times N_2}$,
where the input image has dimension $N_1 \times N_2$, and where
$N_{\alpha, j_{\max}}$ is the total number of $\alpha$-shearlet filters
used by the transform.

The \lstinline{indices} property of the \lstinline{trafo}
object (see line $\# 3$)
can be used to determine to which $\alpha$-shearlet filter
the individual components of the \lstinline{coeff} array are associated.
The value \lstinline{-1} represents the low-pass filter,
while a tuple of the form \lstinline{(j, l, c)} represents the shearlet filter
$\psi_{j,l,\iota}^{(\alpha)}$ as in Equation~\eqref{defn:AlphaShearlets},
where $\iota = 0$ if \lstinline{c} is \lstinline{'h'}
(which stands for the \emph{horizontal} frequency cone),
and where $\iota = 1$ if \lstinline{c} is \lstinline{'v'}
(\emph{vertical} frequency cone).

To explain the differences between the fully sampled and the subsampled
transform, in line $\# 5$, we set up a subsampled transform object
\lstinline{trafo2}. The only difference to the construction of the
\lstinline{trafo} object is that we pass \lstinline{subsampled=True},
and \lstinline{real=False}. The reason for this second change is that---at least
with the current implementation---the subsampled transform can only be used
with complex-valued shearlet filters.
We then compute the coefficients (see line $\# 6$) just as for the
fully sampled transform. Note, however, that the coefficients for the
fully sampled transform were a single 3-dimensional NumPy array.
For the subsampled transform, however, the coefficients are a list of
2-dimensional NumPy arrays.
The reason for this is that the number of coefficients varies from scale to
scale for the subsampled transform.

The \lstinline{indices} property (see line $\# 7$) for the subsampled transform
also differs from that of the fully sampled transform.
The reason for this is that we use \emph{complex} shearlets; therefore, the
frequency plane is divided into four cones (top, or \lstinline{'t'};
right, or \lstinline{'r'}; bottom, or \lstinline{'b'}; and left, or \lstinline{'l'}),
instead of the two cones that are used for real-valued shearlet filters.

The main advantage of the subsampled transform is revealed in line $\# 8$:
The \emph{redundancy} (that is, the number of $\alpha$-shearlet coefficients
divided by the number of pixels of the input image) for the subsampled transform
is much lower, which leads to a lower memory consumption and faster computation times.
While the advantage of the subsampled transform might not be overwhelming in the
given example, it gets more pronounced if one uses a larger number of scales.
For instance, if we use four scales instead of three, then the redundancy of
the fully sampled transform is $41$, while that of the subsampled transform
is only $\approx 11.4$.

\end{appendix}
\end{document}